\documentclass[titlepage,12pt]{article}
\usepackage{graphicx}

\begin{document}
%
%
\def\etal {et al.}
\def\ie {i.\,e.}
\def\etseq {{\em et seq.}}
\def\vs {{it vs.}}
\def\perse {{it per se}}
\def\adhoc {{\em ad hoc}}
\def\eg {e.\,g.}
\def\etc {etc.}
\def\ccpers {\hbox{${\rm cm}^3{\rm s}^{-1}$}}
\def\percc {\hbox{${\rm cm}^3{\rm s}^{-1}$}}
\def\vlsr {\hbox{${v_{\rm LSR}}$}}
\def\vel {\hbox{${v_{\rm LSR}}$}}
\def\vhel {\hbox{${v_{\rm HEL}}$}}
\def\delv {\hbox{$\Delta v_{1/2}$}}
\def\dvel {\hbox{$\Delta v_{1/2}$}}
\def\TL {$T_{\rm L}$}
\def\TC {$T_{\rm c}$}
\def\TEX {$T_{\rm ex}$}
\def\TMB {$T_{\rm MB}$}
\def\TKIN {$T_{\rm kin}$}
\def\TREC {$T_{\rm rec}$}
\def\TSYS {$T_{\rm sys}$}
\def\TVIB {$T_{\rm vib}$}
\def\TROT {$T_{\rm rot}$}
\def\TDUST {$T_{\rm d}$}
\def\TASTAR {$T_{\rm A}^{*}$}
\def\TVIBST {$T_{\rm vib}^*$}
\def\H0 {$H_{\rm o}$}
\def\mic {$\mu\hbox{m}$}
\def\micro {\mu\hbox{m}}
\def\SDOZ {\hbox{$S_{12\mu \rm m}$}}
\def\STWE {\hbox{$S_{25\mu \rm m}$}}
\def\SSIX {\hbox{$S_{60\mu \rm m}$}}
\def\SHUN {\hbox{$S_{100\mu \rm m}$}}
\def\solmass {\hbox{M$_{\odot}$}}
\def\solm {\hbox{M$_{\odot}$}}
\def\solum {\hbox{L$_{\odot}$}}
\def\irlum {\hbox{$L_{\rm IR}$}}
\def\ohlum {\hbox{$L_{\rm OH}$}}
\def\blum {\hbox{$L_{\rm B}$}}
\def\numd {\hbox{$n\,({\rm H}_2$)}}
\def\rhounit {$\hbox{M}_\odot\,\hbox{pc}^{-3}$}
\def\kms {\hbox{${\rm km\,s}^{-1}$}}
\def\kmsyr {\hbox{${\rm km\,s}^{-1}\,{\rm yr}^{-1}$}}
\def\kmsmpc {\hbox{${\rm km\,s}^{-1}\,{\rm Mpc}^{-1}$}}
\def\Kkms {\hbox{${\rm K\,km\,s}^{-1}$}}
\def\percc {$\hbox{{\rm cm}}^{-3}$}    
\def\cmsq  {$\hbox{{\rm cm}}^{-2}$}    
\def\percmsq  {$\hbox{{\rm cm}}^{-2}$}    
\def\persqcm  {$\hbox{{\rm cm}}^{-2}$}    
\def\cmsix  {$\hbox{{\rm cm}}^{-6}$}  
\def\arcsec {\hbox{$^{\prime\prime}$}}
\def\arcmin {\hbox{$^{\prime}$}}
\def\DEGR {\hbox{$^{\circ}$}}
\def\FDEGR {\hbox{$\,.\!\!^{\circ}$}}
\def\ffam {\hbox{$\,.\!\!^{\prime}$}}
\def\ffas {\hbox{$\,.\!\!^{\prime\prime}$}}
\def\ffM {\hbox{$\,.\!\!\!^{\rm M}$}}
\def\ffm {\hbox{$\,.\!\!\!^{\rm m}$}}
\def\HI  {\hbox{HI}}
\def\HII {\hbox{HII}}
%
%
\def \AL {$\alpha $}    
\def \BE {$\beta $}     
\def \GA {$\gamma $}    
\def \DE {$\delta $}    
\def \EP {$\epsilon $}  
\def \alde {($\Delta \alpha ,\Delta \delta $)}
\def \MU {$\mu $}       
\def \TAU {$\tau $}     
\def \tapp {$\tau _{\rm app}$}
\def \tuns {$\tau _{\rm uns}$}
\def \RH {\hbox{$R_{\rm H}$}}         
\def \RT {\hbox{$R_{\rm \tau}$}}      
\def \BN  {\hbox{$b_{\rm n}$}}        
\def \BETAN {\hbox{$\beta _n$}}       
\def \TE {\hbox{$T_{\rm e}$}}         
\def \NE {\hbox{$N_{\rm e}$}}         
\begin{titlepage}
\title{Galactic polarization surveys}
\author{Wolfgang Reich\thanks{email:wreich@mpifr-bonn.mpg.de}
 \\ Max-Planck-Institut f\"ur Radioastronomie
 \\ Auf dem H\"ugel 69, D-53121 Bonn, Germany}
\maketitle

\end{titlepage}

\section{Abstract}

Following the detection of polarized diffuse Galactic emission in 1962 a number
of surveys were undertaken at low frequencies in the following years resulting
in important insights on the local magnetic field, polarization of the giant
radio loops and other Galactic structures, as well as on the properties of the diffuse
magnetized interstellar medium. This field of research experienced a revival
in the eighties and nineties by a number of high resolution observations at low and
high frequencies, which showed a large variety of polarization structures having no
corresponding signature in the total intensity images. 'Canals' and 'Faraday screens'
were reported, which clearly indicate that Faraday rotation in the magneto-ionic
medium may largely vary on small scales. These findings called for a systematic approach
and a number of new unbiased polarization surveys were started. Also new attempts for
absolute calibration are under way, which is a critical issue when interpreting
polarization structures.
This paper reviews polarization survey projects and also summarizes recent results
and interpretations of this rather active field of research.

\section{Introduction}

Galactic radio emission is made up from individual sources like
supernova remnants (SNRs) or HII-regions, which are in their majority distributed along the
Galactic plane and also from diffuse emission originating in the interstellar
medium and extending towards high Galactic latitudes. Diffuse Galactic emission
has a large volume filling factor and consists of synchrotron emission and
the emission from ionized low density gas. Galactic radio sources are highly concentrated
in the thin disk of the Galaxy and studies to derive their physical properties
in most cases require high angular resolution observations for a wide range
of radio frequencies. Polarization observations of synchrotron emitting sources like
SNRs
reveal information about the regularity of the magnetic field, its orientation
within the object and also of the ambient interstellar field. Polarized
radio emission suffers from Faraday rotation within the emitting volume and along
the line-of-sight, which again requires multi-frequency observations to solve
for that. All these effects are valid for diffuse emission as well, where the situation
may be even more complicated by multiple emission layers along the line of sight,
what requires to model the radiation transport conditions in some detail. Studies
of the polarized diffuse Galactic emission received high interest during the
last decade as Faraday rotation effects in the interstellar medium are thought
to be responsible for a zoo of unusual polarized structures, which have no counterpart
in total intensity. The analysis of polarized emission it is also expected to provide
a deeper insight into the composition and structure of the components of the interstellar
medium.

The basis for studies of Galactic polarized emission are large-scale surveys. These
are time consuming projects and a number of surveys are actually running
or planned to map and to analyse these puzzling phenomena in more detail.
They are believed to provide some key information on Galactic magnetic field
 properties and also on weak thermal emission features. These investigations
in some way complement polarization studies of nearby galaxies, which reveal the
   magnetic field structures on scales of a few hundred parsec and larger, which
   are for obvious reasons much more difficult to perform for the Milky Way. A high
    interest in the polarized Galactic emission comes from groups aiming
     to study polarized fluctuations of the
Cosmic Microwave Background (CMB) in the near future, where the weakness of the effect
requires to take the Galactic foreground emission and its statistical properties into account.

\section{History}

After the detection of polarized emission from strong radio sources by
Mayer et al. [1] subsequently also polarized emission from the diffuse Galactic
emission was detected
nearly simultaneously by Westerhout et al. [2] and Wielebinski et al. [3] at 408~MHz.
Those observations finally established the synchrotron emission process as the
major component
of the diffuse Galactic emission at low radio frequencies. However, the percentage
polarization of these observations was much lower than the theoretical maximum, which is close to
70\% for a regular magnetic field. Faraday rotation originating within
the interstellar magneto-ionic medium (MIM) distributed along the line of sight
changes the direction of the original polarization angle, which is orientated along
the electric field vector and orthogonal to the magnetic field direction. Faraday rotation
also causes depolarization, which is rather significant at low frequencies.
Due to the technical possibilities during the sixties observations of extended Galactic
polarization were limited to low frequencies. Furthermore the large beamwidths of
low frequency observations reduce the intensity of the observed polarized signal by vector
averaging effects ('beam depolarization') in addition.

Following its detection in 1962 systematic surveys of diffuse Galactic emission
were carried out for
the northern and southern hemispheres. In the northern sky observations at Cambridge
(7.5-m 'W\"urzburg Riese' dish) by Wielebinski \& Shakeshaft [4] and at Leiden (Dwingeloo
25-m telescope) by Berkhuijsen \& Brouw [5] gave the first insight into the distribution
of polarized emission across the sky. The observing frequency
of 408~MHz means a low angular resolution, which is about $2\DEGR$ for the Dwingeloo telescope
and $7\FDEGR5$ for the Cambridge dish. Southern sky observations by
Mathewson et al. [6] using the Parkes 64-m telescope had a higher angular resolution,
but were severely undersampled. During the following years polarization surveys
were extended towards higher frequencies, e.g. at 610~MHz by the group in
Dwingeloo [7, 8] and at 1407~MHz by Bingham [9] for the northern sky. The
southern sky was observed at 408~MHz and 620~MHz with some complementary data
at 1410~MHz by Mathewson et al. [10].

It is important to note that the definition of the polarized intensity scale
differed among the early observations. Berkhuijsen [11] made a thorough analysis
of the scaling of all available polarization surveys at that time and calculated
the correction factors relative to the definition adopted by the IAU in 1973,
which is based on Stokes parameters.

\section{Galactic synchrotron emission}

The theory of the synchrotron emission process [e.g. 12] describes the radiation
from high energy cosmic-ray electrons in a magnetic field. For a power law
distribution of the energy of the cosmic-ray electrons it can be shown that the
observed emission I$(\nu)$ depends on the number of electrons in the emitting volume
$N_{\rm e}$, the magnetic field component perpendicular to the line of sight
$B_{\bot}$ and the electron energy power law index $\gamma$.

\begin{equation}
  I(\nu) \sim N_{e} B^{(\gamma+1)/2}_{\bot} \nu^{-(\gamma-1)/2}
\end{equation}

thus $\gamma$ is directly related to the observed spectral index
$\alpha = -(\gamma-1)/2$ for observed intensities. Extended emission is the main
component of large-scale surveys and is usually quoted
in units of brightness temperature T$_{b} \sim \nu^{\beta}$,
where the spectral index  $\beta$ is related to $\alpha$ by $\beta = \alpha -2$.

The Galaxy hosts large-scale magnetic fields with a typical strength of a few
$\mu$G. Relativistic cosmic-ray electrons with typical energies between several
hundred MeV and a few GeV are required for measurable synchrotron emission
in the radio range. Diffuse Galactic synchrotron emission dominates the sky
at low frequencies. It is highly concentrated towards the Galactic plane, but
also extends to high latitudes. For instance the high latitude minimum temperatures
observed across the sky at 1.4~GHz are around 0.4~K T$_{b}$.

Discrete synchrotron emitting sources are mainly identified as the remnants from
supernova explosions, which form either expanding shells or diffuse nebula
powered by the wind of the neutron star left from the supernova event. The
shells of SNRs harbour strong magnetic fields, which were swept
up during their expansion. These shock fronts are commonly considered as the
favourable sites for effective cosmic-ray acceleration. After these particle escape
the shock they suffer energy losses mainly by synchrotron cooling, but their
lifetime is long enough that they may diffuse far out off the plane
reaching large heights of a kpc or more.

Synchrotron emission is highly polarized. The degree of polarization p depends on
the energy spectral index of the cosmic-ray electrons and calculates to:
p = $(\gamma+1)/(\gamma+7/3)$. For typical values of $\gamma$ between 2 and 3 the
intrinsic percentage polarization $\rm p_{int}$ calculates between 69\% and 75\%.
These values are higher than those observed, which is due to depolarization effects
(see Sect. 7.3) and also to the degree of regularity of the magnetic field. The
total magnetic field consists of a
random ($\rm B_{ran}$) and a regular component ($\rm B_{reg}$), which give the
total field strength $\rm B_{tot} = (B_{ran}^{2} + B_{reg}^{2})^{1/2}$. Without
depolarization the measured degree of polarization $\rm p_{obs}$ calculates to
$\rm p_{obs} =  p_{int} (B_{reg}^{2}/B_{tot}^{2})$.

\subsection{Radio continuum surveys}

A number of all-sky radio continuum surveys exist in the frequency range up
to 1.4~GHz with angular resolutions of $0\FDEGR6$ at best [13 and references
therein]. Figure~1 shows the all-sky 1.4~GHz survey in total intensity.
An additional large-scale survey is available at 2.3~GHz [14].
At higher frequencies (up to 10~GHz) ground based surveys were restricted to
the narrow band of the Galactic plane, where higher angular resolutions
are needed to resolve the complex emission structures (see Sect. 6.4).
Recently new possibilities were opened by high frequency observations from
satellites, which complement ground based all-sky surveys limited at a few GHz.
These new all-sky maps are from WMAP and cover the frequency range above
22.8~GHz [15] at angular resolutions of about $0\FDEGR5$ or better at
the higher frequencies. A summary of early and more recent total intensity surveys
was given by Wielebinski [16].

The absolute calibration of the high frequency data, which were carried out with
parabolic antennas, is obtained by comparing the survey maps with low angular
resolution sky-horn measurements, which are mostly restricted to scans at
fixed declinations. At low frequencies the Galactic sky temperature is so high that
an absolute calibration can be done by temperature standards.

The spectrum of the diffuse Galactic radio emission varies with frequency and
sky position. At low frequencies the synchrotron emission clearly dominates, while at
higher frequencies the increasing fraction of thermal emission in particular in the
Galactic plane results in a flattening of the spectra. Synchrotron spectra are also
not the same throughout the Galaxy, as they depend on the magnetic field strength and
the cosmic-ray electron spectra, which are known to steepen towards higher energies.
In general flatter spectra are seen at low frequencies.

\begin{figure}
\centering
\includegraphics[bb = 60 22 499 773,angle=270,width=14.3cm,clip=]{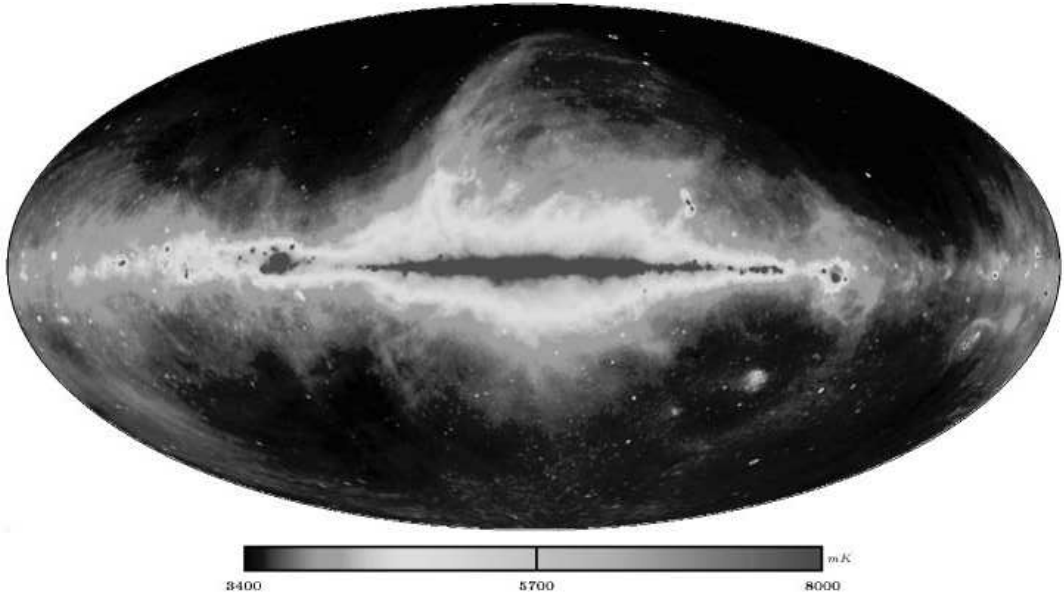}
\caption{Absolutely calibrated 1420~MHz all-sky map in Galactic coordinates
combined from northern sky data
observed with the Stockert 25-m telescope (Bonn University/Germany) [17, 18] and the
southern sky data observed with the Villa Elisa 30-m telescope (IAR/Argentina) [19].
The Galactic Centre is at the centre of the map.
The combined all-sky map [20] includes the isotropic 2.73~K component from the
cosmic microwave background.}
  \label{PR}
\end{figure}

\subsection{The Galactic magnetic field}

There are several methods to obtain information about the global structure of the
Galactic magnetic field and the local field. Our position inside the Galaxy
requires models to get its global structure, of course, there is a strong guidance
by results obtained for nearby galaxies, which can be studied either as face-on or
edge-on objects giving information about the distribution of the disk fields or the field
in the halo. Several reviews of the magnetic field structure in nearby normal
galaxies are available [e.g. 21, 22].

Attempts to model the Galaxy were made by Beuermann et al. [23] and Phillipps et al.
[24], who deconvolved the 408~MHz all-sky survey of Haslam et al. [25].
At 408~MHz synchrotron emission clearly dominates Galactic emission. The emissivity distribution
in the Galaxy was obtained by assuming a regular magnetic
field along a logarithmic spiral structure. The ratio of regular to random magnetic field
of the best model fit was found to be about unity. Strong et al. [26] used a different method based
on comparing $\gamma$-ray emissivity with synchrotron emission to get the magnetic field
strength as a function of the Galacto-centric radius. A steady decrease from the inner Galaxy
to larger radii was obtained. The magnetic field strength close to the Sun is about
6$\mu$G.  Similar results were derived by Berkhuijsen (private communication, see [16,
Fig.~14 ]) assuming equipartition between magnetic field energy and cosmic-ray
energy densities using emissivities from the Beuermann et al. model.

The direction of the local magnetic field was derived using rotation measures (RM)
(see Sect. 5.2) of pulsars of known distance [27, 28, 29, 30]. The direction derived
points towards the Cygnus region in the northern hemisphere and towards the Vela complex
in the southern hemisphere, supporting the view of a general orientation
of the magnetic field along the spiral arms with a pitch angle of a few degrees.
The regular component of the local magnetic field is about 1.4~$\mu$G, just a
small fraction of the total local field of about 6~$\mu$G.
Also for more distant spiral arms the magnetic field direction can be derived using
pulsar RM data. Han et al. [30] derived 4.4~$\mu$G for the regular magnetic field in
the inner Norma arm, where the total field strength on average is about 10~$\mu$G [26].

Of high interest is the existence or
non-existence of "magnetic field reversals", which are predicted under certain conditions by
the dynamo theory (see [31] for a recent review). For nearby spiral galaxies indications for
field reversals are very rare [21]. At present there is a common agreement on a field reversal
between the local Orion arm and the Sagittarius arm at a distance of a few hundred parsec.
For Galactic spiral arms at larger distances possibly more
field reversals might exist. Because of limited data this is discussed controversial.
Significantly larger RM data-sets are needed to decide on this important question.

\section{The interstellar magneto-ionic medium}

\subsection{Thermal emission}

Emission from gas clouds being ionized by the photons of OB-stars is referred to as
thermal emission. These HII-regions are also optically visible depending on the amount
of extinction by dust. Beside discrete gas complexes also a large amount of diffuse
low density gas is present. HII-regions and diffuse gas are concentrated
in the Galactic plane with a smaller scale height compared to synchrotron emission. However,
sensitive H$\alpha$ surveys reveal the existence of thermal gas also far out of the
Galactic plane. Its excitation is not entirely clear yet. Optically-thin
emission spectra, which are seen mostly at frequencies above a few hundred MHz,
are flatter than typical synchrotron spectra. For this reason the fraction of thermal
emission increases towards higher frequencies.

The dispersion measure (DM) from pulsar observations as measured by the time difference of the
pulse arrival time at different frequencies is directly related to the column density of thermal
electrons along the line of sight. Using independent pulsar distances, e.g. as determined
by HI absorption measurements,
the distribution of the thermal electron densities can be derived. On this basis the thermal
electron distribution for the Galaxy was modeled by Taylor and Cordes [32]
and  Cordes and Lazio [33].

The warm ionized gas has typical temperatures of a few thousand Kelvin, but also cold low
density thermal gas seems to exists, which preferably surrounds HII-regions as inferred
from low-frequency absorption measurements [e.g. 34] or recombination line
measurements [35]. Optical
observations indicate an increase of the gas temperature with distance from the Galactic
plane [36], which is also indicated for some edge-on galaxies [37].

\subsection{Faraday rotation}

The coexistence of magnetic fields and thermal gas in the interstellar medium results
in Faraday rotation. The amount of Faraday rotation also called the 'Faraday depth' is
(in the simplest case) equivalent to the rotation measure (RM), which is defined
as the slope of the polarization angle $\phi$ versus $\lambda^{2}$,
and calculates for a line of sight L [pc]:

\begin{equation}
  RM [rad~m^{-2}] = 0.81 \int_L n_{e} [cm^{-3}]~B_{\|} [\mu G]~dL[pc]
\end{equation}

with $n_{e}$ being the thermal electron density and $B_{\|}$ the magnetic field component along
the line of sight. A detailed discussion of 'Faraday depth' related issues was recently made
by Brentjens and de Bruyn [38].

RM is positive in case the magnetic field direction points towards the observer and vice versa.
The amount of Faraday rotation depends on the observing wavelength, where
the observed polarization angle $\phi$ depends on the intrinsic polarization angle $\phi_{o}$
and RM by $\phi$ = $\phi_{o}$ + RM~$\lambda^{2}$. Observations at two frequencies need to
be combined to calculate RM. However, there is an ambiguity in RM, which can be solved by
adding observations at a third frequency. Narrow-band polarimetry around a certain frequency
is the preferred observing technique as the beams are very similar, depolarization variations
are small and therefore the emission originates from the same volume. However, the low
 signal-to-noise ratio of narrow-band polarimetry may limit the accuracy of the measurements,
 that small RM differences are not easy to measure. A new method called 'Faraday Rotation
Measure Synthesis' was recently introduced [38], which allows
to analyse contributions from multiple sources along the line of sight and improves the
signal-to-noise ratio of narrow-band polarimetry. An application
of this impressive new technique to multi-channel wide-field Westerbork polarization
images was recently made for observations of the Perseus cluster [39].

In the case of a uniform mixture of thermal emission and synchrotron emission the so called
'slab-model' [40] describes  the observed emission as a function of frequency. However,
the distribution of emission components in the interstellar medium is certainly less uniform.
Emitting sources with internal and external Faraday rotation are observed through the diffuse
magneto-ionic medium along the line of sight, where clouds, shells or bubbles with no or
very little total intensity emission ('Faraday screens') are embedded.

\section{Galactic polarization surveys}

\subsection{Observing technique}

Galactic synchrotron emission is linearly polarized. This requires to measure the
Stokes parameter U and Q, from which the polarized intensity PI and polarization
angle $\phi$ calculate as

\begin{equation}
  PI = \sqrt{(U^{2} + Q^{2})}
\end{equation}

\begin{equation}
  \phi = 0.5~atan~(U/Q)
\end{equation}

Early polarization measurements were made with single-dish telescopes using
 rotating crossed dipoles as feeds.  The power difference between the dipoles was
measured. Special care must be taken for the antenna pattern ellipticity [4].
Another method was to measure the cross-correlation
between the two feed dipoles [2]. There are two possibilities to
obtain a set of Stokes parameters from correlation. In case the feeds
couple out linear polarization Stokes parameters I, Q and V are obtained
after correlation and a $90\DEGR$ phase shift of one signal.
Coupling out two circular polarization components gives Stokes I, Q and U after
correlation, which is the preferred configuration to measure linear polarization
needed for Galactic polarization work.

The feeds and their adjustment in the telescope are critical components for
successful polarization measurements. At low frequencies dipoles are used
and at higher frequencies corrugated waveguide feeds.
A number of methods are used to change linear polarization components
into circular components and various microwave components are used for this
purpose.

\begin{figure}
\centering
\includegraphics[bb=76 49 500 744,angle=90,width=14cm,clip=]{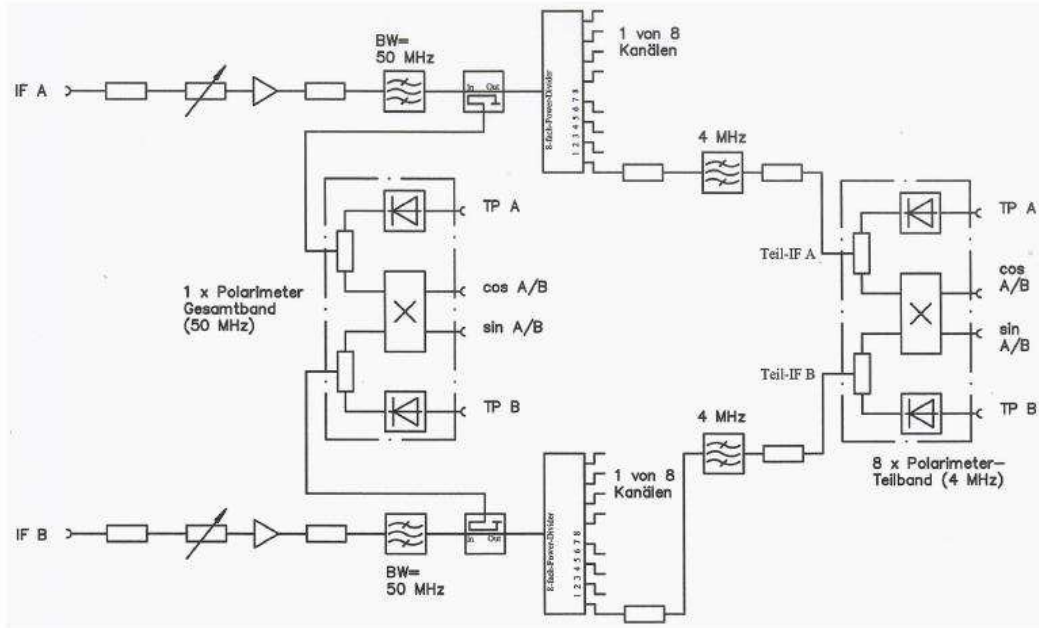}
\caption{Block diagram of an IF-polarimeter with 8 narrow and one wide
channel, which is attached
to the L-band receiver installed at the Effelsberg 100-m telescope (courtesy
of O. Lochner).}
  \label{pol}
\end{figure}

Figure~\ref{pol} shows, as an example, a 8-channel narrow-band IF-polarimeter
and an additional broad-band channel attached to the L-band receiver at the Effelsberg
100-m telescope. It is used for 1.4~GHz and 1.7~GHz polarization observations and for RM
determination. At other telescopes correlator backends have been installed
which allow to measure simultaneously wide bands split into hundreds of narrow
channels. Analysis of these polarization data cubes require new techniques as the
already mentioned 'Faraday Rotation Measure Synthesis' [38], for instance.
At frequencies of 5~GHz and higher analog IF-polarimeter with a bandwidth of several
hundred MHz up to a few~GHz are available for measurements of the usually very weak polarized
signals at these frequencies. This is possible since depolarization across the
band becomes less important at high frequencies (see Sect. 7.3).

\subsection{Instrumental effects}

For polarization measurements a number of instrumental effects must be taken
into account. The antenna feed and other frontend components cause losses
or cross-talk between the polarization channels. The polarimeter response, in addition,
may not be circular and the observed U and Q signals depend on the polarization angle
or the parallactic angle for telescopes with an AZ-EL-mount. Instrumental effects
may also vary with time. All these influences must be calibrated in a suitable way using
measurements of unpolarized radio sources and an analysis of the polarized calibration
signal to find the instrumental parameters. With the measured
instrumental terms a 4x4 correction matrix, the 'Mueller matrix',
is set up, which simultaneously transforms the observed polarization
components into the four Stokes parameters.

\begin{figure}
\centering
\includegraphics[bb=69 142 492 587,angle=270,width=8.5cm,clip=]{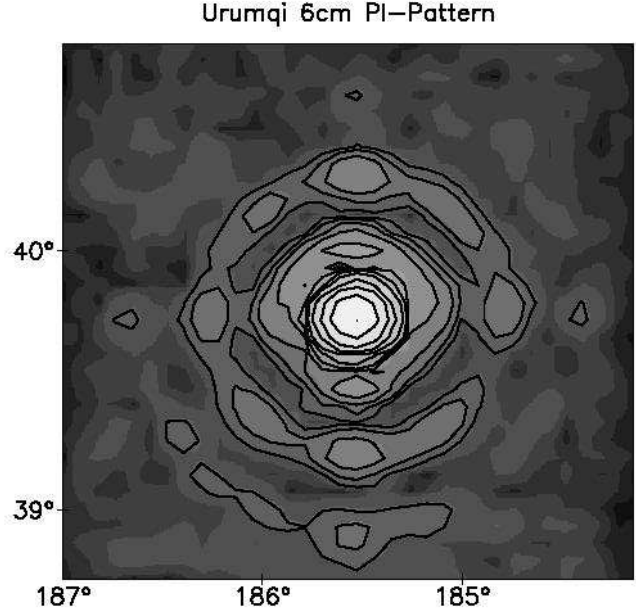}
\caption{6~cm antenna pattern for polarized intensities of the Urumqi 25-m
telescope using the signal received from the INSAT-3B satellite. Contours run
in steps of 3~dB down to -30~dB. The field size is  $2\FDEGR2 \times 2\FDEGR1$.
The sidelobes are enhanced towards north, east, south and west due to the
influence of the four telescope's support legs.}
  \label{urum}
\end{figure}

\begin{figure}
\centering
\includegraphics[bb=126 64 528 652,angle=270,width=10.5cm,clip=]{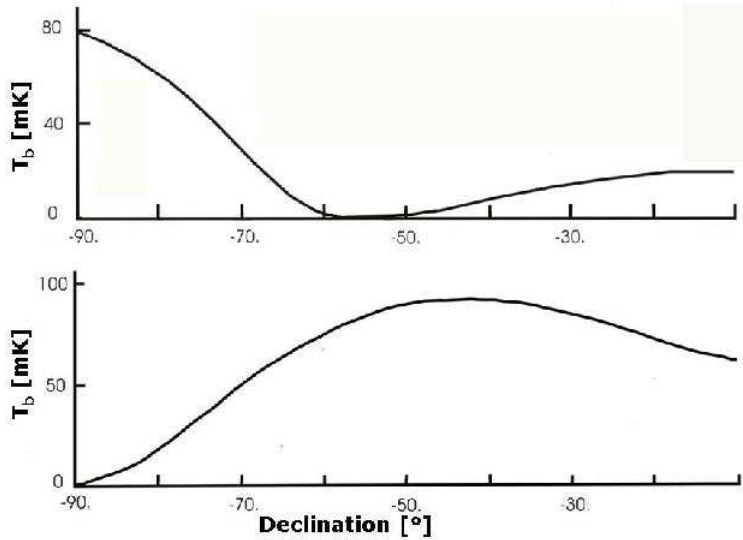}
\caption{Ground radiation of the Villa Elisa 30-m telescope at 1.4~GHz.
The profiles were derived from averaging a large number of scans along declination from
various sky directions of the two polarimeter channels. The zenith for the Villa Elisa
telescope corresponds to a declination of about $-55\FDEGR15$. The intensity scale is
in mK $\rm T_{b}$ with an arbitrary zero-level. [Testori et al., in prep.]}
  \label{ground}
\end{figure}

\begin{figure}
\centering
\includegraphics[bb=20 15 284 266,width=8cm,clip]{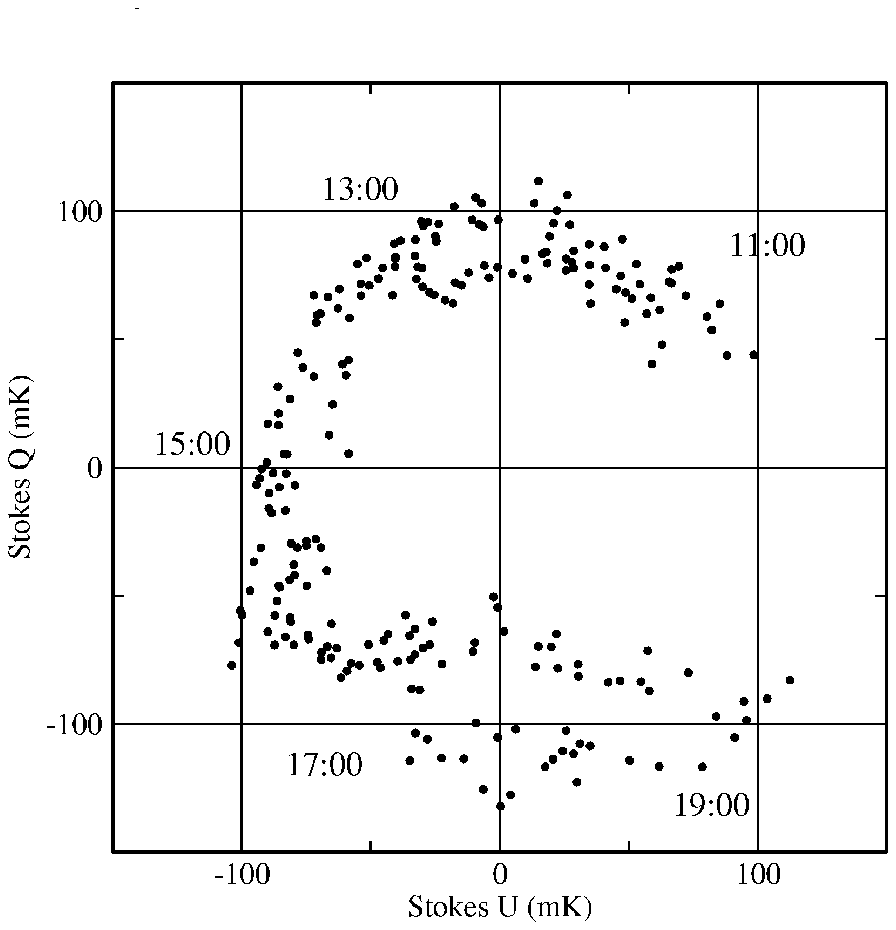}
\caption{Example of an absolute polarization measurement (with preliminary calibration)
towards the NCP at 1.4~GHz made with the 26-m DRAO telescope [43]. Apparent sky rotation
causes a systematic variation of the signals in the 'U' and 'Q'-channel of the polarimeter
with time (as indicated). Polarimeter offsets and the contribution from the ground are
required to be constant during these measurements.}
  \label{mw-ncp}
\end{figure}

\begin{figure}
\centering
\includegraphics[bb=340 125 630 555,width=6cm,clip]{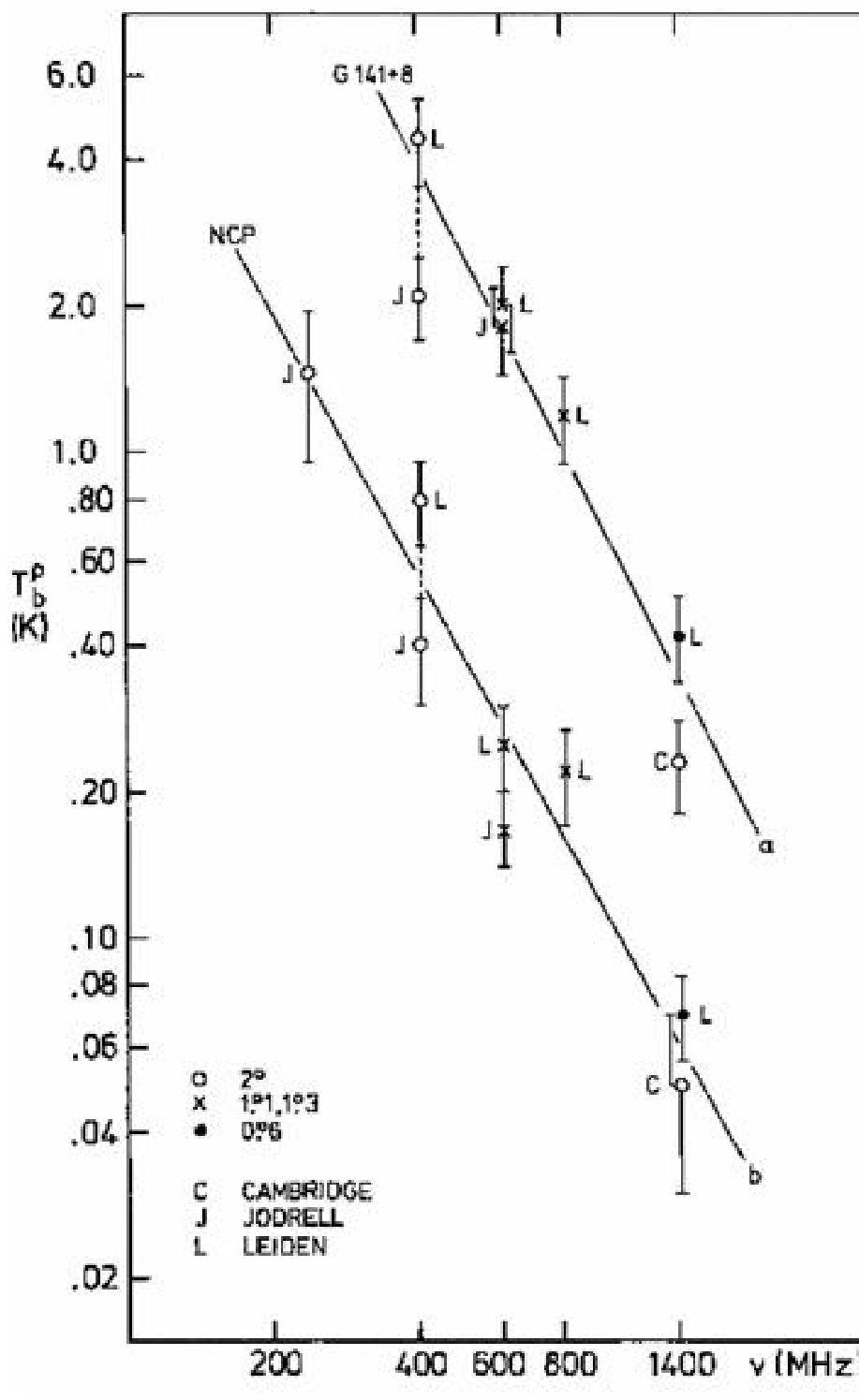}
\caption{Low frequency polarization spectra towards the North Celestial Pole (NCP)
and the calibration point at $l,b = 141\DEGR,8\DEGR$ [11]. The polarized intensity spectra
are significantly flatter than those of the total intensity synchrotron emission.
For the NCP data shown the spectral index is $\beta$ = -1.8. Other NCP observations
confirm the flat spectrum: $\beta$ = -1.87$\pm 0.05$ [41]
or $\beta$ = -2.06$\pm 0.1$ [42].}
  \label{elly}
\end{figure}

Sidelobes caused by strong polarized signals are similar in structure and level
compared to the total intensity sidelobes. An example of a 6~cm polarization
pattern is shown in Fig.~\ref{urum}. Unpolarized sources
often show a 'butterfly'-shaped antenna response for the Stokes parameter U and Q
in the area of the main beam. This causes a ring-like shaped characteristic for
the polarized intensity response of an unpolarized source. Far-sidelobes are highly
polarized and this may cause problems in particular for large-scale survey
observations. The sidelobes pick up radiation from the ground and the observed
sum of these spurious signals does not vary necessarily in a systematic way.
In general ground radiation increases
towards lower elevation often rather different in U and Q. An example for
ground radiation profiles derived for the Villa Elisa 1.4~GHz polarization
survey of the southern sky is shown in
Fig.~\ref{ground}. Depending on the site of the telescope (e.g. mountains in the surroundings)
the ground radiation may also vary with azimuth direction. The celestial poles were
often used for absolute
polarization temperature measurements, because all instrumental contributions are
(in principle) constant in this direction and by apparent sky rotation the polarization
components describe a circle in the U,Q-plane as shown in Fig.~\ref{mw-ncp}, whose
radius gives the polarized intensity. The low frequency spectrum for the North Celestial Pole
is included in Fig.~\ref{elly}.

The far-sidelobe structure of a telescope depends on the reflecting or scattering support
structures within the telescope [44]. Although the absolute level of the far-sidelobes
is rather low (typically much below -50~dB) day time observations are often affected or
become impossible due to scattered solar radiation picked up by the far-sidelobes.

\subsection{Available large-scale surveys}

\begin{figure}
\centering
\includegraphics[bb=112 49 485 744,angle=270,width=14cm,clip=]{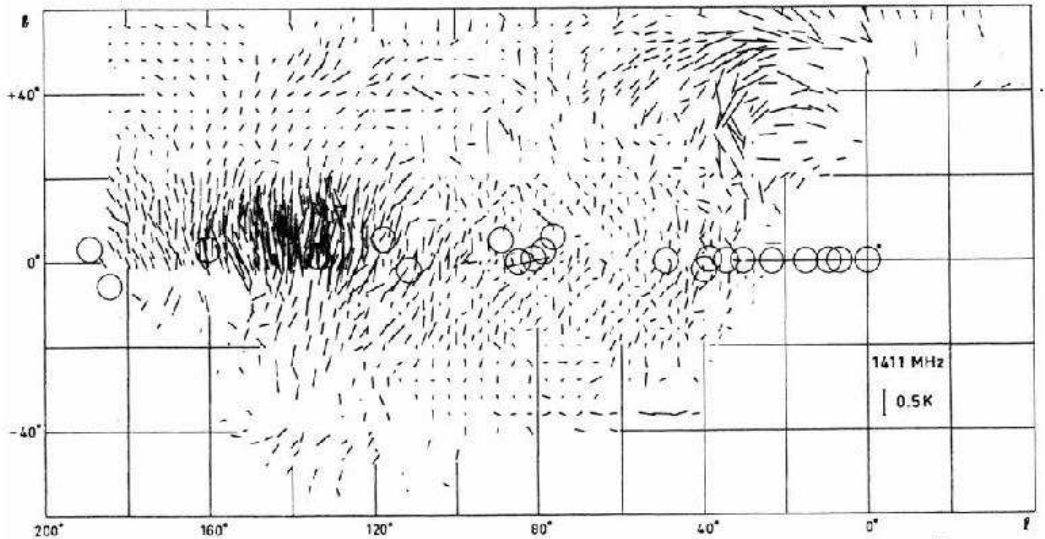}
\caption{The Leiden-Dwingeloo 1.411~GHz polarization survey of the northern sky
showing polarization bars proportional to polarized intensity in E-field direction [45].}
  \label{dwing}
\end{figure}

A series of linear polarization surveys of the northern sky between 408~MHz and 1411~MHz
were carried out using the 25-m Dwingeloo telescope and published in 1976 [45 and
references therein]. All these surveys were on an absolute scale
being corrected for the polarized ground radiation and also for Faraday rotation
occurring in the ionosphere, which needs to be taken into account for frequencies
below about 1.4~GHz. The angular resolution of these surveys varies
according to the wide range of frequencies between $2\DEGR$ at 408~MHz and
$0\FDEGR6$ at 1.411~GHz. The sky was not entirely mapped and the sampling was
not complete. However, the data are available in numerical form and were widely
used until today. Figure~\ref{dwing} shows the result of the 1.411~GHz survey in the form
of polarization bars in E-field direction.
Another series of polarization surveys were carried out with the Jodrell Bank
76-m telescope. A compilation of all large-scale northern and southern sky
Galactic surveys published until 1976 is included in the paper by Spoelstra [46].

More recently several large areas of the sky were mapped with the Westerbork telescope
at 327~MHz [47-50] and more work is in progress. These low frequency maps have
arcminute angular resolution and reveal a wealth of small-scale polarization
structures. The Westerbork observations were recorded in several narrow channels,
what allows RM determination. The results of these observations have been interpreted
in a series of papers [48-55].

As seen from Fig.~\ref{dwing} the Leiden-Dwingeloo northern sky survey at 1.411~GHz has no
complete coverage and is also severely undersampled in large areas of the sky. A much
more densely sampled new survey
at 1.4~GHz was recently carried out using the 26-m telescope at the DRAO/Canada [56]
at about the same angular resolution of $36\arcmin$, but five times higher sensitivity
compared to the Leiden-Dwingeloo survey. The DRAO survey results from declination
scans with the telescope's position fixed in the meridian. By apparent sky rotation fully
sampled data along right ascension
were obtained. In declination the sampling of the data varies between $0\FDEGR25$
and about $2\FDEGR5$ and therefore some interpolation is needed. The survey is described
by Wolleben et al. [56], more details are given in [43, 57].
The DRAO data were tied to the absolutely calibrated Leiden-Dwingeloo data to find
their temperature offsets. Polarization data from the Effelsberg 100-m telescope were used to
fix the temperature scale. The data from the DRAO polarization survey
are available at $http://www.mpifr-bonn.mpg.de/survey.html$ (for
selected fields in different projections) or
at $http://www.drao.nrc.ca/26msurvey$ or $http://www.mpifr-bonn.mpg.de/div/konti/26msurvey$
(for the entire set of survey data).

\begin{figure}
\centering
\includegraphics[bb=133 149 456 645,angle=270,width=14.3cm,clip=]{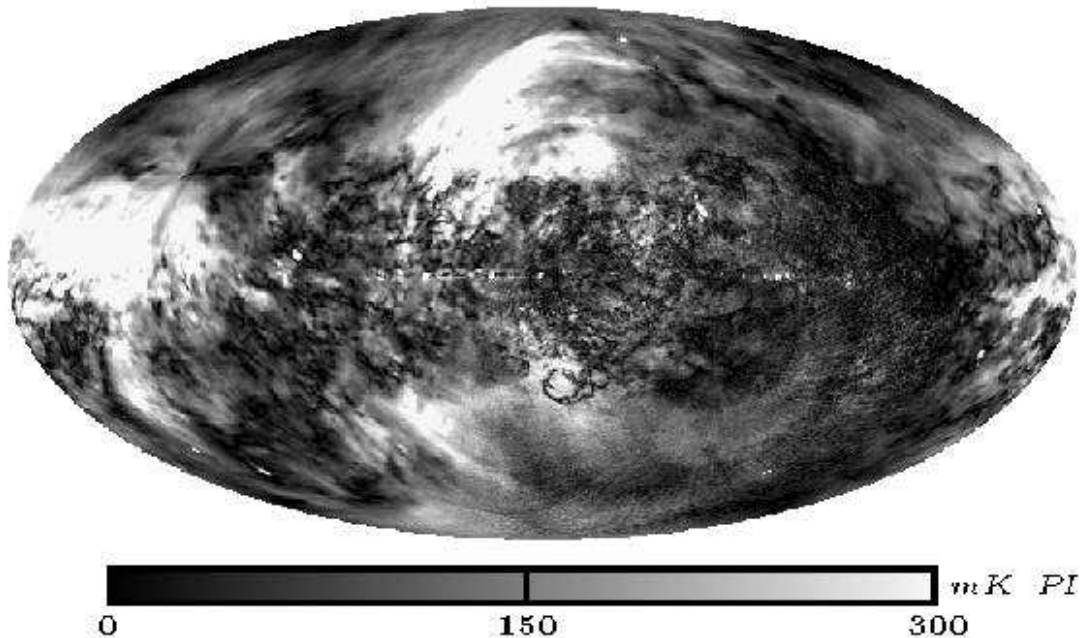}
\caption{All-sky 1.4~GHz map in polarized intensity obtained by combining the
northern sky polarization survey obtained with the DRAO 26-m telescope [56] and
southern sky polarization data from the Villa Elisa 30-m telescope [58] after
adjustment. The angular resolution of the map is $36\arcmin$, however, the northern sky
data are not fully sampled in declination and thus were interpolated [56].}
  \label{pi}
\end{figure}

Corresponding southern sky polarization data at 1.4~GHz will be also available next time. They
were observed with the Villa Elisa 30-m telescope simultaneously with the total intensity
southern sky survey [19]. Some reduction procedures and first maps were already
presented by Testori et al. [58]. Unfortunately no absolutely calibrated data are
available for the southern sky so far. However, there is a large region of overlap
between declination $-10\DEGR$ and $-28\DEGR$ with the DRAO northern sky survey, whose data
are used to find the absolute polarization level for the southern sky.
Villa Elisa maps of the strong, highly polarized and sufficiently extended radiogalaxy
Centaurus A were used for scale and angle calibration in respect to maps from the Parkes 64-m
telescope. A combination of the northern sky and the southern sky survey is shown in
Fig.~\ref{pi}, which is the first all-sky polarization map obtained so far.

\subsection{Galactic plane surveys}

\subsubsection{Radio continuum surveys}

Radio continuum surveys of the Galactic plane need higher angular resolutions
than all-sky surveys to resolve the large number of individual sources in the thin disk of the
Galaxy from the diffuse emission, which is most intense in the Galactic plane as well. These
surveys form the basis for detailed investigations of individual objects, but also
for spectral investigations. Numerous
surveys were made with single-dish telescopes. In particular the Effelsberg 100-m telescope
was used for northern sky observations at 1.4~GHz [59, 60], 2.7~GHz [61, 62, 63]
and at 4.9~GHz [64], where the angular resolution is about $2\ffam6$.
The Parkes 64-m telescope
was used for mapping the southern Galactic plane at 5~GHz with $4\ffam1$ angular
resolution [65]. 2.4~GHz Parkes maps were published by Duncan et al. [66].
The highest frequency Galactic plane survey comes from the Nobeyama 45-m telescope
at 10~GHz with $2\ffam7$ angular resolution [67].

Also synthesis telescopes were used to survey the Galactic plane, but they miss most of the
diffuse emission and their maps of extended sources are often incomplete.
Synthesis telescope surveys are, however, very well
suited to measure compact sources in a complex environment. A combination of synthesis telescope
surveys with single-dish surveys is needed to obtain maps including emission on all angular scales.
This is for instance a standard procedure for the 408~MHz and 1420~MHz maps from the
CGPS ('Canadian Galactic Plane Survey')[68], which are either combined with the 408~MHz
all-sky survey [25] or the Effelsberg telescope surveys at 1.4~GHz [59, 60].

\subsubsection{Polarization surveys}

\begin{figure}
\centering
\includegraphics[bb=48 71 492 673,angle=270,width=13.9cm,clip=]{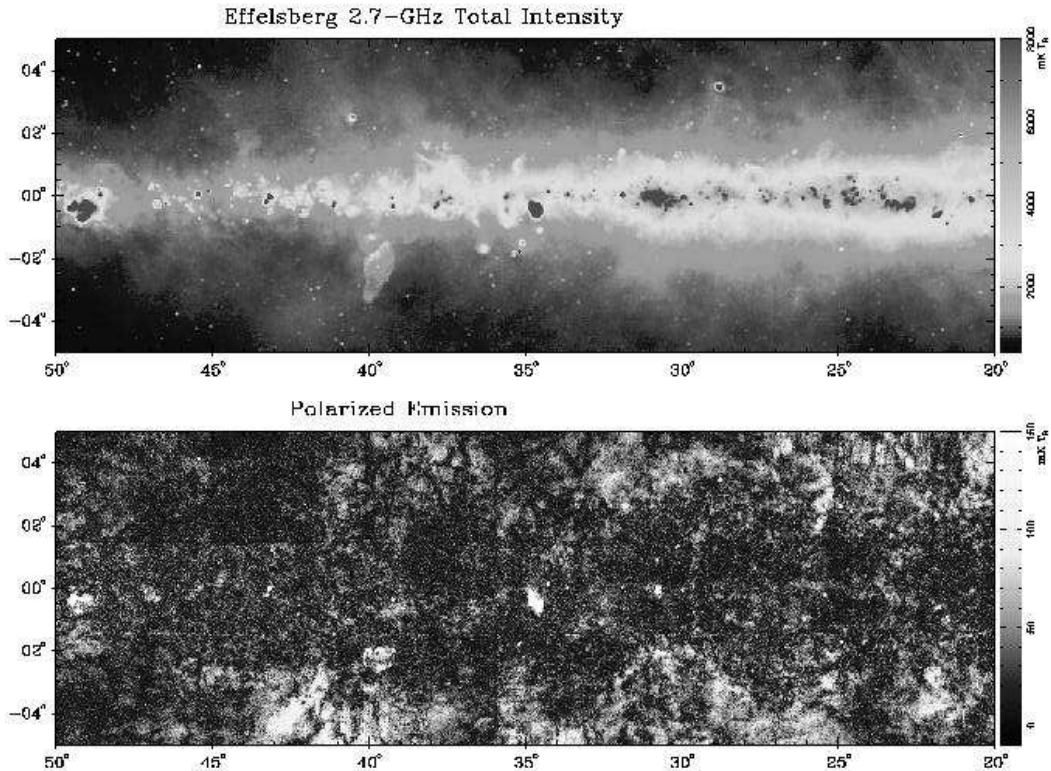}
\caption{Section of the 2.7~GHz Effelsberg survey of the Galactic plane
clearly demonstrating the anti-correlation between total intensities (upper panel)
and polarized intensities (lower panel) [62, 70].}
  \label{11cm}
\end{figure}

The Galactic plane survey at 2.7~GHz carried out with the Effelsberg 100-m telescope at
$4\ffam3$ angular resolution includes simultaneous observations of linear polarization
[69]. The area between $4\FDEGR9$ and $74\DEGR$ Galactic longitude with a latitude
range of $\pm 5\DEGR$ was complete in polarization [70]. The Parkes telescope maps at
2.4~GHz complement the Galactic plane in southern directions [71]. Some discussion and analysis
of the properties
of the polarized emission along the Galactic plane was made by Duncan et al. [70].
A section of the Effelsberg 2.7~GHz survey is shown in Fig.~\ref{11cm}, where
the anti-correlation between total intensity and polarized intensity indicates depolarization
by thermal matter, which is known to have a more narrow latitude distribution compared
to synchrotron emission. The 2.7~GHz continuum and polarization survey data and also a number
of other published radio survey data are freely available
through the web from the MPIfR survey sampler: {\it (http://www.mpifr-bonn.mpg.de/survey.html)}.

\subsection{Running projects}

The observations for the Effelsberg 1.4~GHz 'Medium Galactic Latitude Survey' (EMLS),
which covers the Galactic plane from $25\DEGR$ to $240\DEGR$ in longitude
for latitudes within $20\DEGR$, is in an advanced stage of reduction [72]. Its
observational methods and calibration procedures as well as a first set of example maps
were published by Uyan{\i}ker et al. [73, 74].
The angular resolution of the EMLS is $9\ffam4$. It is confusion limited in total intensity
at a rms-noise level of 15~mK T$_{b}$. The rms-noise in Stokes U and Q is about 8~mK T$_{b}$.
Finally missing large-scale components will be added from the recently completed 26-m
DRAO polarization northern sky survey, which is at an absolute temperature level [56].
Example maps from the EMLS are displayed in Figs.~11, 12 and 20.

The Effelsberg telescope was also used at 1.4~GHz to map a high latitude strip using a
special observing strategy to recover also the large-scale polarization components.
Preliminary results were published by Abidin et al. [75]. Higher structured
U and Q maps when compared to polarized intensity maps indicate the presence of substantial
Faraday rotation at high latitudes even at 1.4~GHz. The fractional polarization
level is up to about 30-40\%.

The Arecibo 1000-ft telescope is planned to be used for a multi-beam L-band survey including
linear polarization
to be carried out by the GALFA Consortium [76]. The frequency from 1225~MHz to 1524~MHz
will be covered simultaneously at about $3\arcmin$ angular resolution. A polarimeter with
1000~channels will be used. According to the sky accessible with the
Arecibo telescope the observations will include sections of the Galactic plane, but also
high latitude regions of the Galaxy.

Higher angular resolution polarization surveys at 1.4~GHz are underway from the DRAO
synthesis radio telescope in the context of the 'International Galactic Plane Survey' (IGPS), previous CGPS ('Canadian Galactic Plane Survey')[68]. The IGPS (CGPS) project includes
continuum, polarization, HI and CO observations at about $1\arcmin$ angular resolution
and provides other complementary data as an unique basis for a wide range of Galactic studies.
The 1.4~GHz polarization data are observed in four 7.5~MHz wide channels placed on either side of
a central band, from which only HI images are derived. This allows the determination of RMs.
An example polarization map is shown in Fig.~17. The Cygnus section from the polarization
survey was published including an analysis of a variety of polarization structures
by Uyan{\i}ker et al. [77]. Other polarization data from the CGPS were discussed elsewhere
[e.g. 78-80].

The ATCA synthesis telescope was used to carry out the 'Southern Galactic Plane Survey'
(SGPS) in continuum, linear polarization and HI at about the same angular resolution as
the DRAO survey.
The polarization data were collected in 12 bands, each 8~MHz wide, in the frequency range
from 1332~MHz to 1436~MHz. A first section of the polarization survey was discussed by
Gaensler et al. [81], up-dated information and more results from the SGPS are shown by
Haverkorn et al. [82]. It is planned to complement the SGPS for missing large-scale
components by additional observations with the Parkes 64-m telescope.
Also the Villa Elisa southern sky polarization survey (Sect. 6.3) will be available to add
the largest structures.

As already mentioned in Sect. 6.3 also new Westerbork polarization observations between
300~MHz and 400~MHz of large high Galactic latitude fields are in progress.

\begin{figure}[htb]
\centering
\includegraphics[bb=55 57 550 723,angle=270,width=14.3cm,clip=]{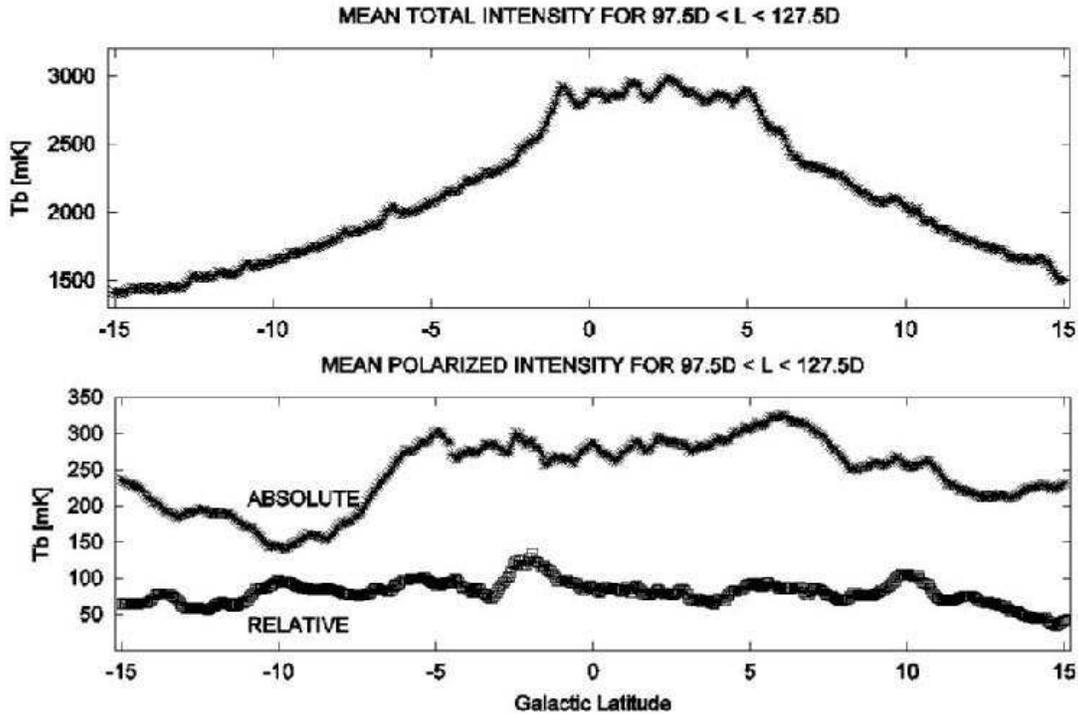}
\caption{Averaged 1.4~GHz intensities from the EMLS for a section of the Galactic plane in
the second quadrant shown as a function of Galactic latitude. Average polarized intensities
are shown for the EMLS ('Relative') and
after adding missing large-scale components from the Leiden-Dwingeloo survey ('Absolute',
see Sect. 7.1).}
  \label{EMLS-AV}
\end{figure}

There are clear indications that polarization surveys available so far
up to 1.4~GHz are mostly tracing emission structures of local origin. Comparing
the all-sky total intensities (Fig.~\ref{PR}) and polarized intensities (Fig.~\ref{pi})
clear depolarization effects are indicated by the patchy and almost constant distribution
of polarized emission for absolute latitudes below about $30\DEGR$  for the inner Galaxy.
Depolarization of local origin is also obvious from Fig.~\ref{11cm} and
Fig.~\ref{EMLS-AV}. The distance ('polarization horizon' [78]) where entire depolarization
along the line of sight occurs depends on the properties of the interstellar medium
and also on the angular resolution of the observations.
In general emission from much larger distances can be traced at high latitudes
than close to the Galactic plane. Clearly observations at higher frequencies than 1.4~GHz
or 2.7~GHz are needed. Currently the 25-m telescope at Nanshan station (Urumqi Observatory
NAOC/China) is engaged in a 5~GHz polarization survey, which is a common NAOC/MPIfR project.
At that frequency the distance range to observe polarized structures is about ten times
larger than at 1.4~GHz. The intention of the Urumqi observations are to
map a similar area as covered by the 1.4~GHz EMLS survey. The angular
resolution of $9\ffam5$ will be about the same for both surveys.
To reach high sensitivities a re-build Effelsberg 5~GHz receiver and
a broad-band polarimeter were installed at the Urumqi telescope. A brief description of
the system was already given by Sun et al. [83].
The survey will be sensitive enough to trace polarized emission structures below
1~mK $\rm T_{b}$. Recently also the Torun 30-m telescope is used for polarization
observations of sections of the Galactic plane at 4.7~GHz using a MPIfR polarimeter [84].

The new 5~GHz polarization data need absolute calibration, which is not easy to perform at such
high frequencies because of the weakness of the polarized signal in relation to
systematic effects by the instrument and the environment.
Efforts in this direction are undertaken at the Urumqi 25-m telescope.
Alternatively some modelling of the high frequency signals based on absolute 1.4~GHz survey data
[56] in combination with RM data from the Leiden-Dwingeloo surveys [45] might be used.
These RM data, however, are derived from measurements between 408~MHz and 1.411~GHz, which
is not ideal. Therefore it is planned
to continue with polarization survey observations at the 26-m DRAO telescope in 2006 using a
new multi-channel polarimeter to carry out the 'DRAO/MPIfR RM-Survey' for the northern
sky above $-30\DEGR$ declination and about $30\arcmin$ angular resolution with full sampling.

\section{Analysis of polarized emission}

\subsection{Absolute calibration}

The much more structured polarized emission when compared to the total
intensity distribution calls for an explanation in terms  of physical
parameters of the magneto-ionic interstellar medium and numerous attempts
have been already made so far. However, here must be a warning: all the
new synthesis telescope surveys miss large-scale structures depending on the
smallest baselines used.  At very low frequencies the missing U and Q structures
are likely small because of strong RM dispersion. Also surveys
from large single-dish telescopes suffer from missing large-scale structures
exceeding the size of the mapped region, although
these scales are normally much larger than for synthesis telescopes. Structural
interpretations of total intensities are
rather little affected by these missing components, since a temperature
offset or emission with a gradient across the feature of interest does
usually not change its morphology. In polarization,
however, the missing emission is a vector and when vectors are added
the effect on the small-scale structures may be rather significant.
For instance, enhanced small-scale emission may turn into a depression feature when
adding large-scale emission or vice versa.

Several methods exist which combine low resolution
with high resolution data. In case of the EMLS the combination with the
Dwingeloo absolute polarization data was described by Uyan{\i}ker et al. [73]. In brief,
the undersampled Dwingeloo U and Q data were interpolated to a regular grid
and convolved. The corresponding Effelsberg data for the same field were convolved to
the same effective beamwidth. Optionally spatial filtering may be applied to both maps.
The difference to the Dwingeloo data was then interpreted as the missing large-scale
component and is finally added to the Effelsberg data at their original
angular resolution of $9\ffam4$.

Another technique is used when combining synthesis telescope data with single-dish
data, where both maps are merged in the Fourier plane with appropriate weighting of
the spatial frequencies in the range of overlap. In this case the smallest spacing
of the synthesis telescope has to be smaller than the size of the single-dish telescope.

\begin{figure}
\centering
\includegraphics[bb=20 20 538 772,angle=270,width=14cm,clip=]{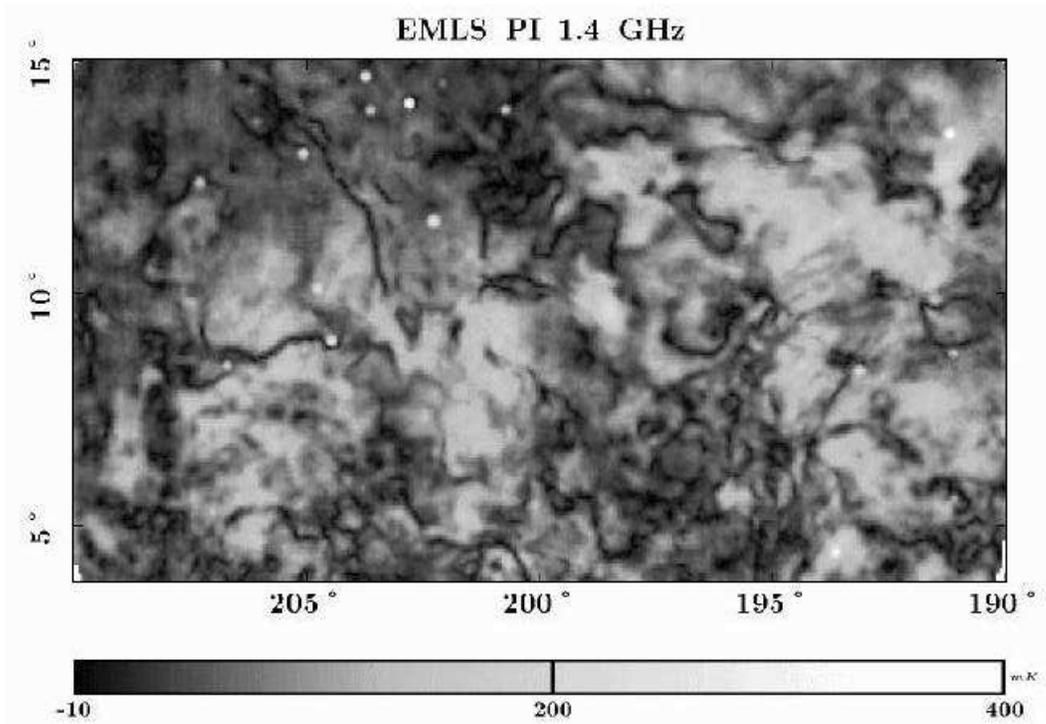}
\caption{Map of 1.4~GHz polarized intensities as observed
with the Effelsberg telescope [74] in a section of the Galactic anti-centre.
The angular resolution is $9\ffam4$ and the rms-noise in Stokes U and Q
is 8~mK $\rm T_{b}$. }
  \label{anti1}
\end{figure}

\begin{figure}
\centering
\includegraphics[bb=20 20 551 772,angle=270,width=14cm,clip=]{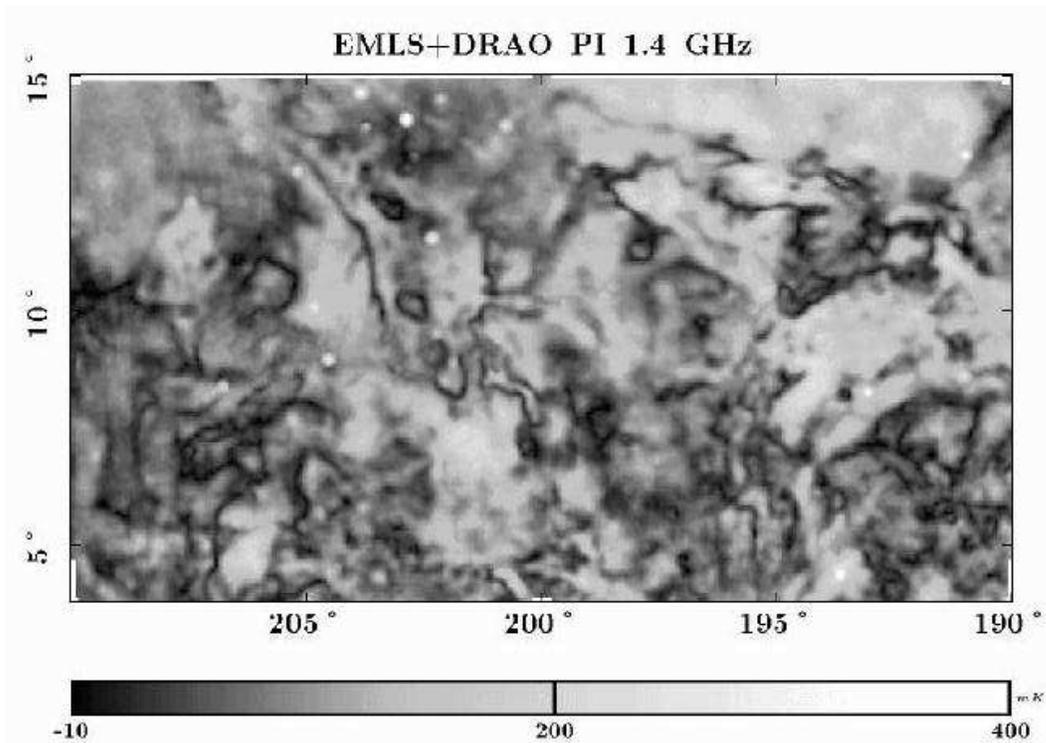}
\caption{Polarization intensity map for the same field as in Fig.~\ref{anti1}, but
including large-scale data from the 26-m DRAO telescope [43].}
  \label{anti2}
\end{figure}

\begin{figure}
\centering
\includegraphics[bb=60 50 555 760,angle=270,width=12cm,clip] {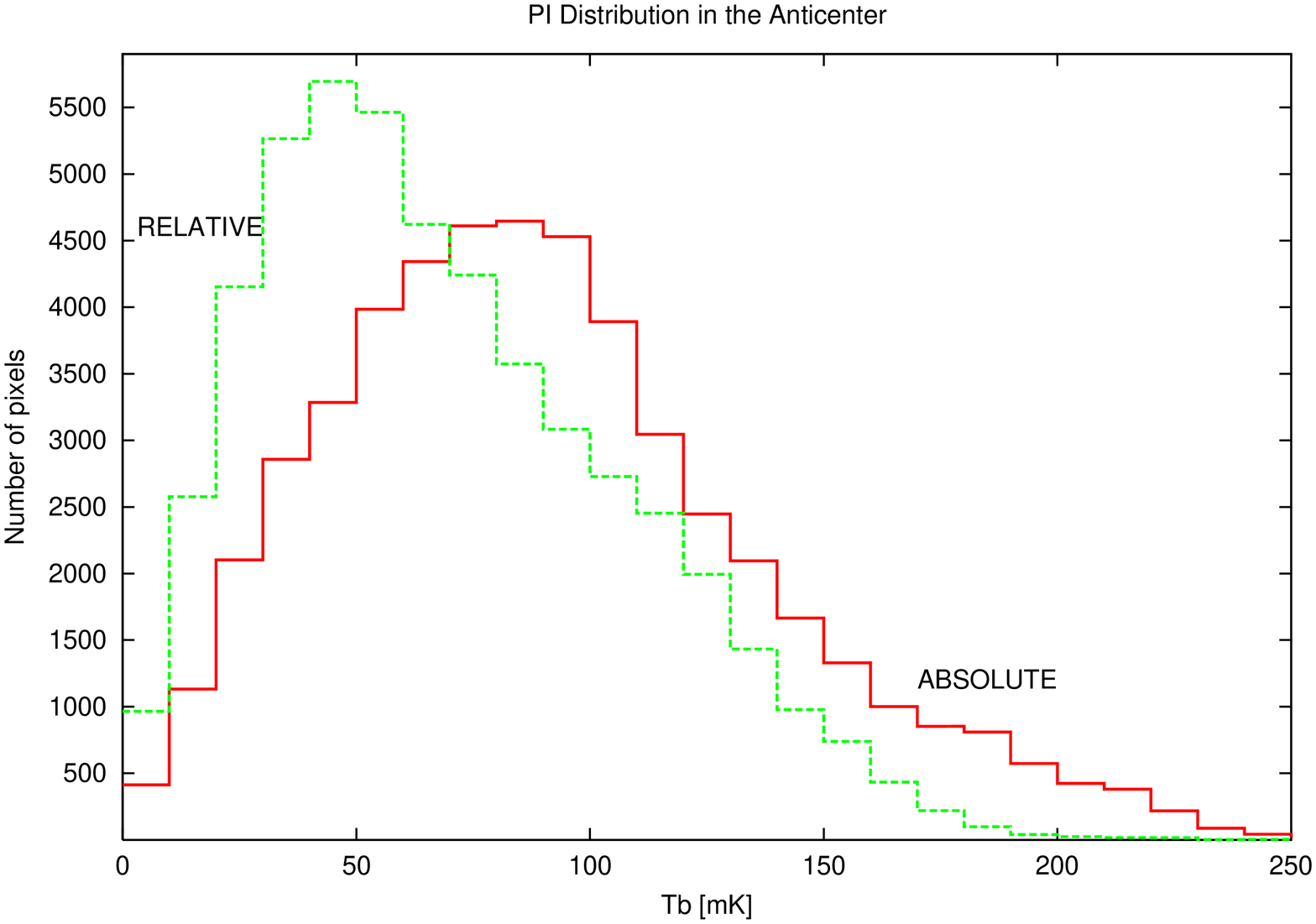}
\caption{The polarized intensity spectrum for the Effelsberg map (Fig.~\ref{anti1}) labeled
'Relative' and the Effelsberg map including large-scale components from the
26-m DRAO survey (Fig.~\ref{anti2}) labeled 'Absolute'.}
  \label{anti-spec}
\end{figure}

As already mentioned, the effect of adding the missing large emission in Stokes U
and Q is non-linear for polarized intensity and the polarization angle (equations 3 and 4)
and may result in significant changes in morphology. Also the
distribution of polarized intensities and polarization angles changes.
In Fig.~\ref{anti1} an example map
from the EMLS is shown, where at the time of publication [74] no absolute polarization
data were available. This field is located in the Galactic anti-centre a few
degrees out of the Galactic plane, where the level of total intensity is quite
low and dominated by extragalactic sources (see [74] for the total intensity image).
Figure~\ref{anti2} shows the same field with large-scale U and Q components from the
26-m DRAO survey [43] added, which clearly causes morphological changes
for most but not all small-scale features. The pixel spectrum of polarized intensities
for both maps is shown in Fig.~\ref{anti-spec} for comparison, where the mean level
of polarized emission differs by about 22~mK or about 2$\times$ the
rms-noise of the original EMLS map.

\begin{figure}
\centering
\includegraphics[bb=20 20 489 772,angle=270,width=14cm,clip=]{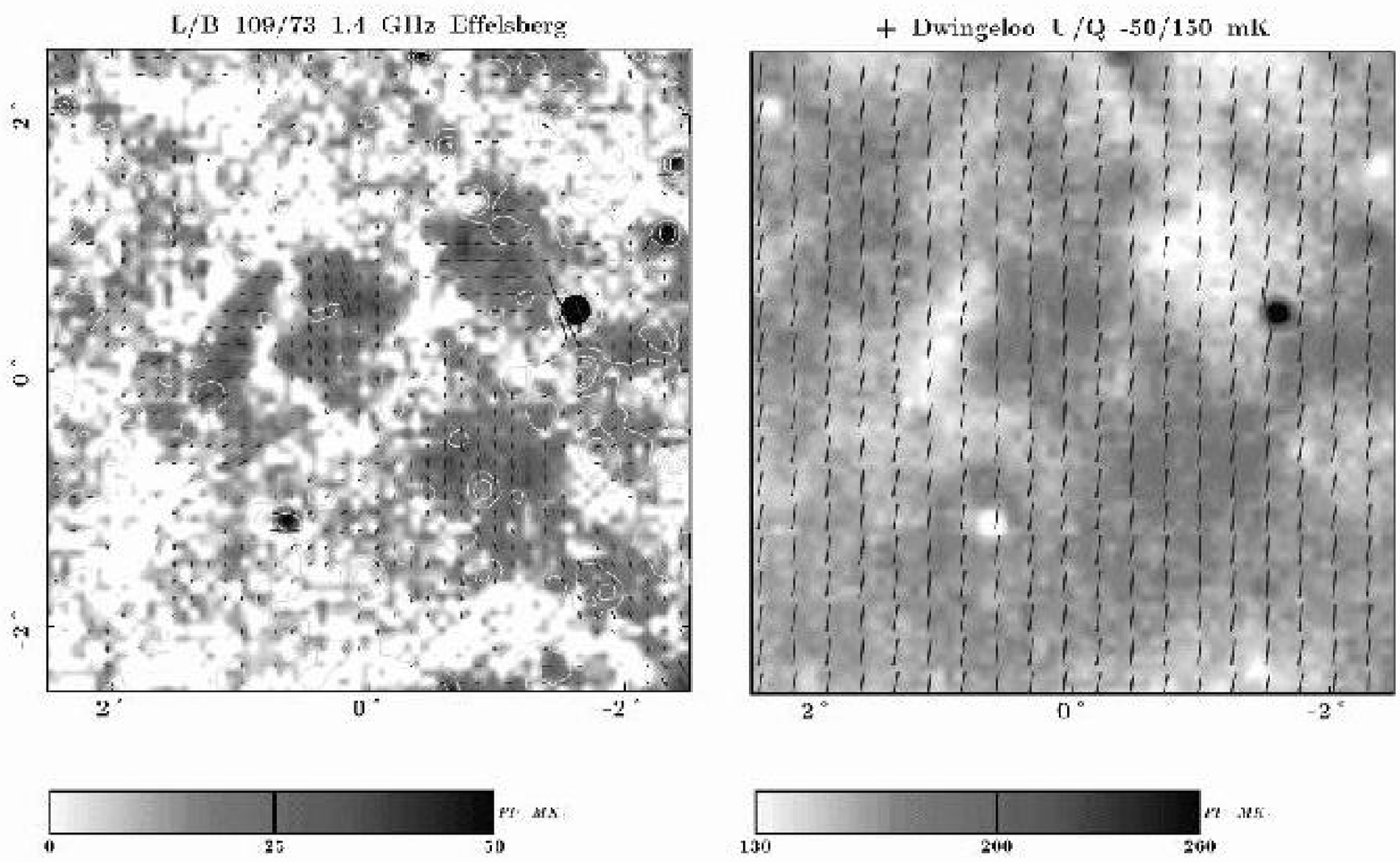}
\caption{High latitude field centered at l,b = $109\DEGR, 73\DEGR$ observed
with the Effelsberg 100-m telescope at 1.4~GHz (left panel). White contours
indicate total intensities. The right panel shows the same map after
addition of large-scale U and Q offsets taken from the Leiden-Dwingeloo survey [45].
Polarization bars are in E-field direction [85].}
  \label{g109-pi-map}
\end{figure}

\begin{figure}
\centering
\includegraphics[angle=270,bb=60 50 555 760,width=12cm,clip] {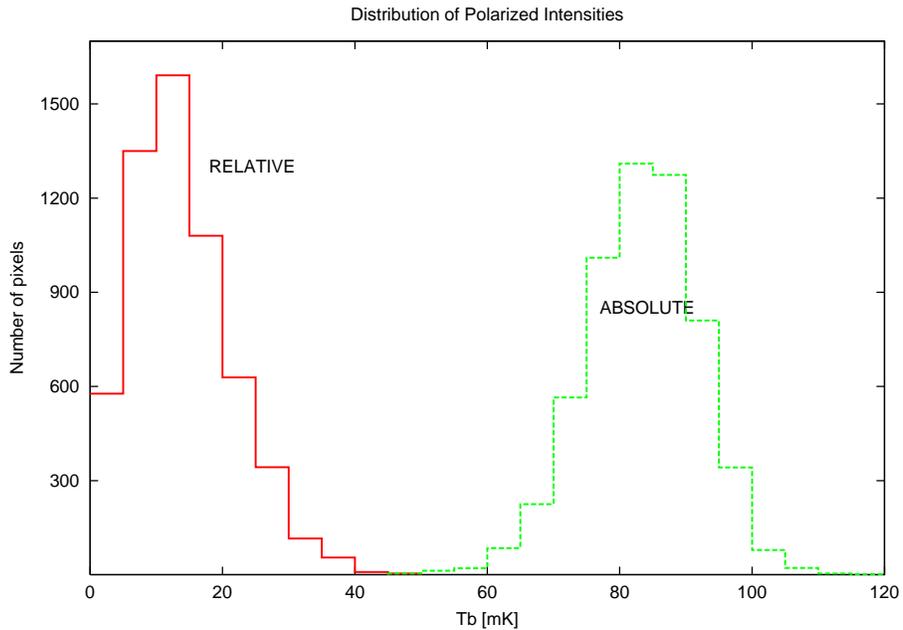}
\caption{Pixel distribution of polarized intensities for the Effelsberg map
 (Fig.~\ref{g109-pi-map}, left panel) labeled 'Relative' and the Effelsberg map
 including large-scale components from the Leiden-Dwingeloo survey [45]
 (Fig.~\ref{g109-pi-map}, right panel) labeled 'Absolute'.}
  \label{g109-pi}
\end{figure}

\begin{figure}
\centering
\includegraphics[angle=270,bb=60 50 555 760,width=12cm,clip] {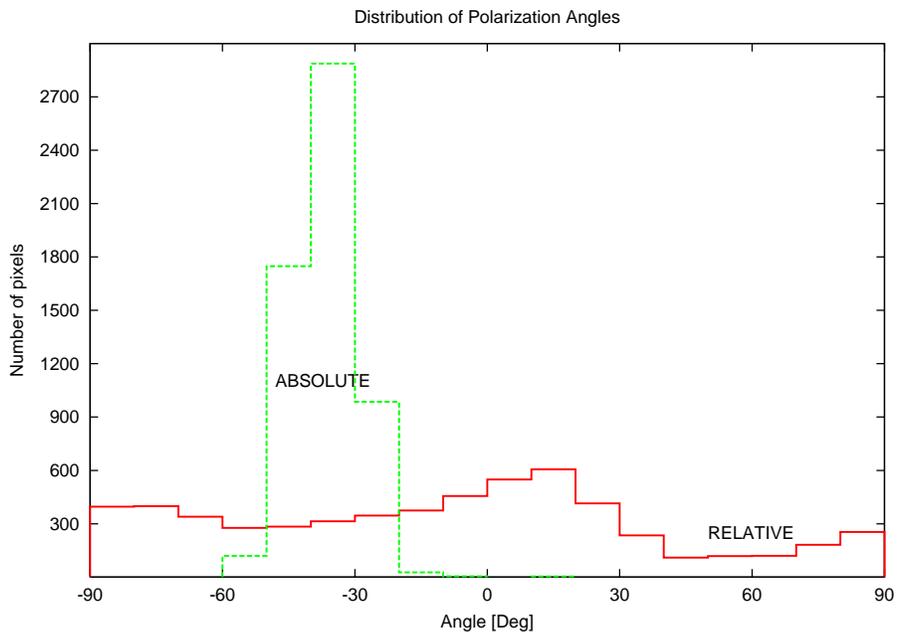}
\caption{Pixel distribution as in Fig.~\ref{g109-pi}, but for the polarization angles.
Adding large offsets in U and Q largely reduces the range of polarization angle variations.}
  \label{g109-ang}
\end{figure}

\begin{figure}
\centering
\includegraphics[bb=20 20 473 772,angle=270,width=14cm,clip=]{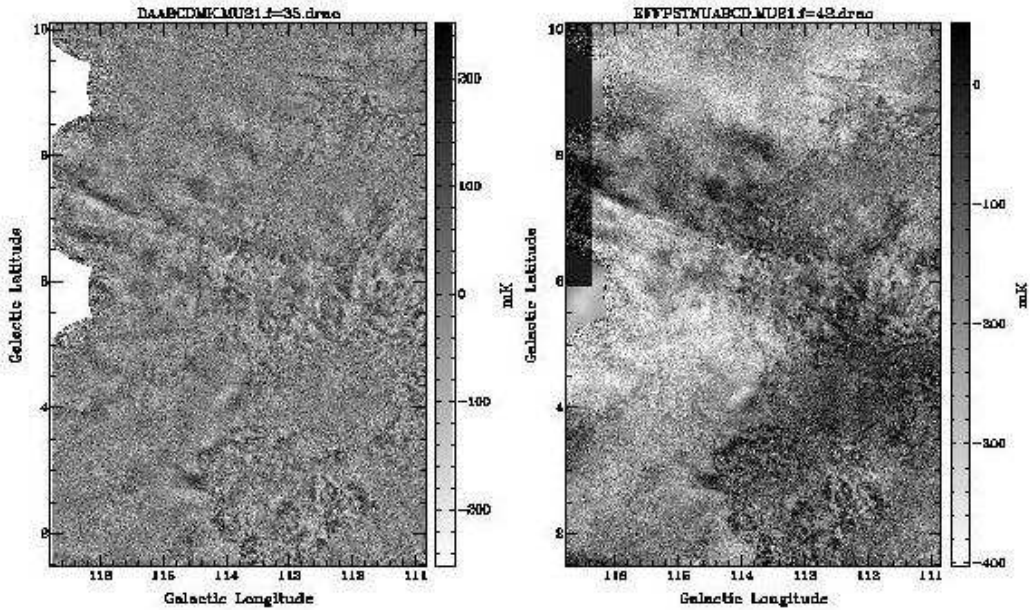}
\caption{A Stokes U image using only DRAO interferometric CGPS data (left panel).
 A preliminary combination of Effelsberg 100-m and DRAO 26-m single-dish data
 with interferometric CGPS data (right panel).
  Note that what appeared to be a filament running North--South at l =
  $114\DEGR$ in the lower half of the left panel is now revealed to be
  a front between two regions.  The upper half of the right panel shows
  another front [86].}
  \label{rob}
\end{figure}

In the case of high latitude fields the effect of missing large-scale structures
causes a dramatic change of the polarization morphology. Figure~\ref{g109-pi-map}
(left panel) shows a $5\DEGR \times 5\DEGR$ field observed at 1.4~GHz
with the Effelsberg telescope in a similar way as the EMLS observations [85].
No extended total intensity
features are seen exceeding a few times the noise level. However, numerous extragalactic
sources dominate the field, a few of them are clearly polarized. The map of polarized
intensities measured with the Effelsberg telescope shows a number of distinct patches
with a typical size of about $0\FDEGR5$. Typical peak polarized intensities are about
25~mK $\rm T_{b}$.
Figure~\ref{g109-pi-map} (right panel) shows the same field after adding the missing
large-scale emission for
this field. From the Leiden-Dwingeloo 1.411~GHz data [45] -50~mK in U  and 100~mK in Q
are estimated and added as constant offsets to the corresponding Effelsberg U and Q maps.
The resulting differences in the polarized intensity maps and in the distribution of
the polarization angles are large. After absolute calibration the polarized emission
patches from the Effelsberg map are now seen either as enhancements of the uniform
background polarization in case their polarization vectors are aligned with those
of the diffuse large-scale emission. Otherwise they cause depressions in the polarized
emission. Therefore the morphology of the polarized intensity distribution is totally
different when the large-scale components are added. Because vectors are added also the
structure function is changed. The distribution
of the polarization angles is now reduced to small variations around a mean value for the
field set by the dominant large-scale U and Q offsets.
The strong changes are clearly reflected in the pixel distributions as displayed in
Figs.~\ref{g109-pi} and \ref{g109-ang}.

From missing associated total intensity features in the Effelsberg map it is already
clear that the polarized patches must be caused by Faraday effects, which in view of
the high Galactic latitude are likely not far distant from us. In general, local Faraday
screens cause a larger observational effect on background polarization than more distant
ones. There must be a difference in the magnetic field orientation of the foreground
and the background emission that after Faraday rotation by a screen both fields are more aligned
and thus the sum of both components exceeds the value seen outside of the screen. Also the
opposite case occurs that by a screen's Faraday rotation the misalignment between foreground
and background is enhanced. In that case a decrease
of the polarized intensity is seen in respect to its surrounding. Multi-frequency
data to calculate RMs are needed to decide whether the differences in Faraday rotation
are due to fluctuations around a certain (large enough) RM value or the magnetic fields
in the Faraday screens have an opposite direction.

An example for the differences in morphology and physical interpretation resulting from
combining EMLS single-dish map with higher resolution CGPS data from the DRAO
synthesis telescope is shown in Fig.\ref{rob}. Some filamentary structures
seen as isolated features in the CGPS map are revealed to mark the boundary between
different extended regions when the single-dish data are included [86].

\subsection{RM determination}

\begin{figure}
\centering
\includegraphics[bb=40 30 278 501,width=4.5cm,clip=]{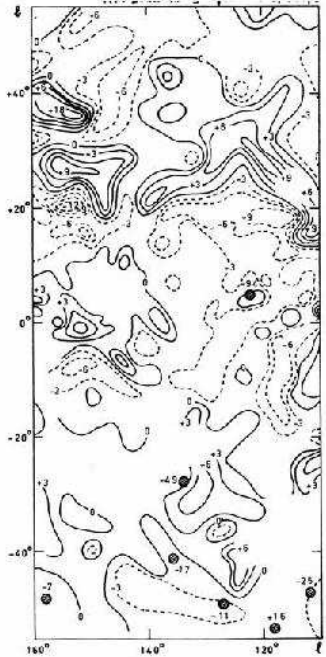}
\caption{RM map based on the Leiden-Dwingeloo polarization surveys covering partly
the Cetus Arc and Loop III [46]. The black dots
indicate RM values for extragalactic sources, which are in general much larger than
the RM values for the diffuse Galactic emission as observed at low frequencies. }
  \label{rm-titus}
\end{figure}

The RM of the diffuse Galactic emission was determined by various authors using
available low frequency polarization surveys [42, 45]. Figure~\ref{rm-titus} shows an
example for a RM--map calculated by Spoelstra [46].
RMs were determined by fitting a linear slope to
the observed polarization angles as a function of $\lambda^{2}$. RM maps from
low frequency surveys show all quite low values, in particular when compared to
the RM values from extragalactic sources in the field, which trace the
Faraday rotation from the entire interstellar medium along the line of sight
through the Galaxy and do almost not suffer from depolarization effects.
The conclusion from this result is that the observed diffuse low frequency
polarized emission is of local origin. Of course, the early RM maps should be
interpreted with some care. A valid RM determination requires that the emission
originates from the same volume, which is questionable when data at 408~MHz
are compared with those at 1.4~GHz, except when they are very local. In addition,
the observation should have the same beamwidth to avoid an effect from polarization
vector averaging across the area of the beam, except that the emission region
has a very uniform distribution of polarization vectors and is larger than all the
beams involved.

Today's RM determination from multi-channel narrow-band polarimeter avoid these
effects. Gaensler et al. [81] using the ATNF compact
array (ATCA) in connection with a multi-channel backend for mapping the polarized
emission of the southern Galactic plane obtain rather high RM values ranging
within about $\rm \pm150~rad~m^{-2}$ in the direction of thermal emission regions.
However, the ATCA data miss the large-scale emission components and
these may well influence the RM values measured and thus their physical interpretation.
A new project to determine the RM distribution of the diffuse polarized
Galactic emission on an absolute scale at L-band, where the beamwidth variations
over the band are small, will start soon at the 26-m DRAO telescope (see Sect. 6.5).

The $\lambda^{2}$ dependence of polarization angles is an important diagnostic tool [see 38],
however, it is destroyed in case of the superposition
of various polarization components with different RM or by depolarization effects or by
missing large-scale structures. From two frequency observations one
can not decide whether the $\lambda^{2}$ dependence is given and thus RMs should be taken
with care. The correct interpretation of observed RMs needs additional modelling efforts,
for instance, to determine the physical parameter of passive
Faraday rotating structures ('Faraday Screens') at a certain distance (see Sect. 7.4).

\subsection{Depolarization effects}

Faraday rotation in the interstellar medium along the line of sight leads to depolarization.
In addition there are instrumental effects
as the beamwidth and the bandwidth of the observation, which also might cause depolarization.
All these effects have been already extensively discussed in the literature and are briefly
summarized here:

{\it Bandwidth~depolarization} occurs when the polarization angles vary across
the frequency band, which reduces the observed amount of polarized emission. The depolarization
DP, which is defined as DP = PC$_{obs}$/PC$_{int}$, the ratio between the observed and the
intrinsic polarization, calculates as DP = sinc (2~RM~$\lambda^{2}~\delta\nu/\nu$), where
$\delta\nu$ is the bandwidth of the observations. For narrow observing bands, high frequencies
or small RM values DP becomes negligible.

{\it Beam~depolarization} occurs in case polarization vectors of different
 orientation are unresolved by the telescope beam. In order to separate
{\it beam~depolarization} from depolarization effects in the interstellar medium
polarization data measured at different wavelengths must be compared at the same
angular resolution.

In his classical paper Burn [40] discussed
differential Faraday rotation (was also called depth depolarization), internal
Faraday dispersion and Faraday dispersion in
an external screen. These basic concepts and formula are most often used in the interpretation
of polarized emission. Of course, it is rather obvious that more complex scenarios than
the uniform distribution of the magneto-ionic components are required to describe recent
 Galactic polarization observations [38]. Sokoloff et al. [87] made an attempt in that
direction and demonstrate that
non-uniform effects have a strong effect on the observed polarization distribution.

Burn's 'slab model' [40] is often used to calculate the {\it internal depolarization}
of a source, where thermal electrons, relativistic electrons and magnetic fields are
uniformly mixed. Then DP calculates as DP = sinc(2~RM~$\lambda^{2}$). For the case of
significant RM fluctuations $\sigma_{RM}$ one gets DP = $|\frac{1-\exp(-S)}{S}|$, where S = $2~\sigma_{RM}^2~\lambda^4 - 4i~\lambda^2~RM$.
{\it External depolarization} depends entirely on $\sigma_{RM}$ in the foreground
medium: DP = $\exp(-2\sigma_{RM}^2\lambda^4)$ (for more details see [87]).

\subsection{Faraday screens}

\begin{figure}
\centering
\includegraphics[bb=174 189 421 653,width=9cm,clip=]{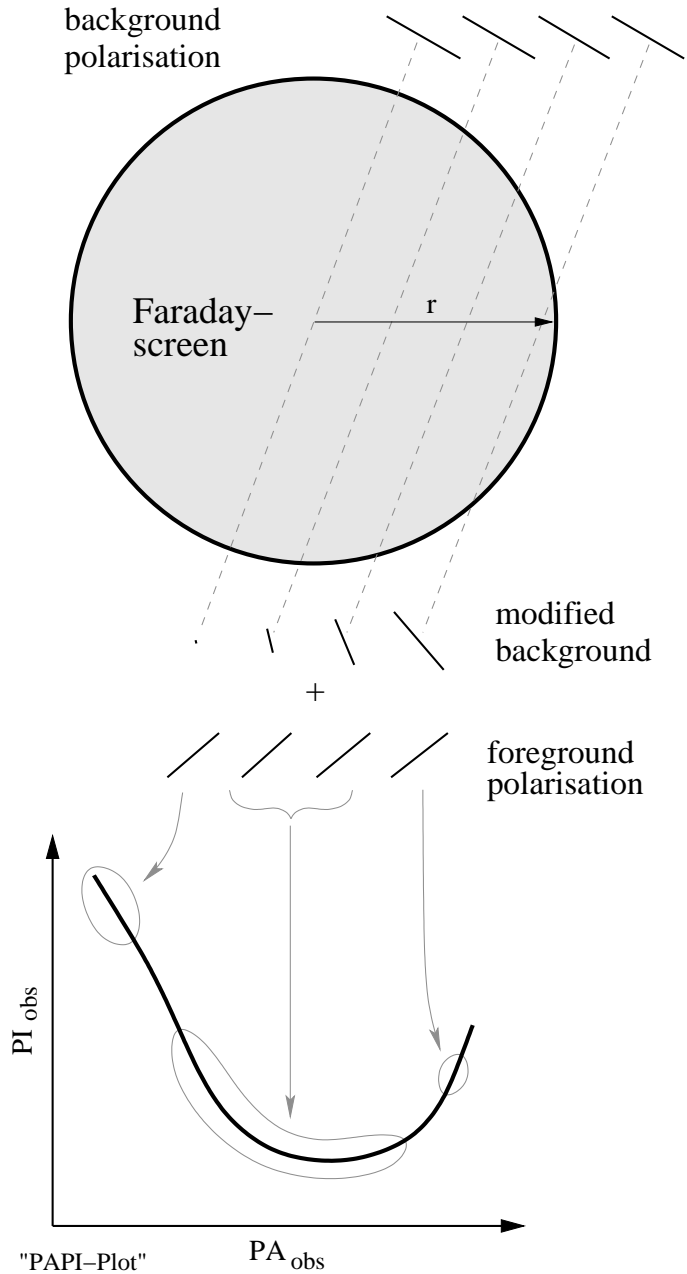}
\caption{A spherical cloud acting as a Faraday screen. The
line of sight length through the cloud is taken as proportional to RM
(and also to DP). The background polarization is modified by the Faraday
screen and adds to the foreground polarization. The observed polarization for
such a region has a characteristic polarization intensity versus polarization
angle dependence (PAPI-plot). Based on this model multi-frequency
polarization data at an absolute level were analysed by Wolleben and Reich [89, 90]
to derive physical parameters of Faraday screens located at the boundaries of
local molecular clouds in the Taurus region. Also the properties
of the Galactic foreground and background polarization can be determined by the
model. }
  \label{model}
\end{figure}

Excessive Faraday rotation in the magneto-ionic interstellar medium may be caused
either by an enhanced thermal electron density or by a stronger or more regular
magnetic field component along the line of sight. For example,
RM values of extragalactic sources may be enhanced when observed in the direction of a
HII-region. Strong magnetic fields are known to exist in the shells
of SNRs as the result of interstellar magnetic field compression by their expanding shock fronts,
which cause a strong increase of the synchrotron emissivity
and in most cases also strong linear polarization. In addition to Faraday rotating sources
with a clear signature in total intensities, the polarization surveys reveal a class
of Faraday screens, which are very weak or invisible (at a certain sensitivity level)
in the corresponding total intensity survey map, but impose clear effects in the polarization
angle distribution and/or the polarization intensity distribution. In some cases weak
H$_{\alpha}$ emission can be seen, but for many Faraday screens available H$_{\alpha}$
surveys seem not to be sensitive enough
to trace them and just allow to derive an upper limit for the thermal electron density.
These Faraday screens have a low electron density and thus need an enhanced regular magnetic
field along the line of sight direction to account for the observed Faraday rotation.
Discrete Faraday screens were first reported and discussed by Gray et al. [88] and Wieringa
et al. [47].

Any interpretation of the observed RMs towards a Faraday screen needs its distance
to get its physical parameters. This is in general not easy to measure. Wolleben and
Reich [89, 90] made observations of Faraday screens located nearly exactly at the boundaries
of some Taurus molecular clouds, for which the distance is known to be 140~pc.
A full analysis of absolutely calibrated data, which are at least needed at two frequencies,
gives information on the physical properties of the Faraday screens and
the fraction of foreground to background polarized emission for its distance.
The observations used in this study are from the Effelsberg 100-m
telescope at 1408~MHz, 1660~MHz and 1713~MHz. Zero-spacings were added from the
1411~MHz Leiden-Dwingeloo survey to the 1408~MHz map. Zero-spacings for the other
two frequencies were extrapolated with a spectral index of $\beta$ = -2.7, which
is close to the total intensity spectral index. It was also assumed that
RM = 0~rad~m$^{-2}$ is valid for the diffuse emission in this area (see Fig.~\ref{rm-titus},[46]).
These assumptions seem quite reasonable, but also
reflect the fact that the availability of absolutely calibrated data and RM information
is rather limited.

\begin{figure}
\centering
\includegraphics[bb=20 20 480 772,angle=270,width=14.3cm,clip=]{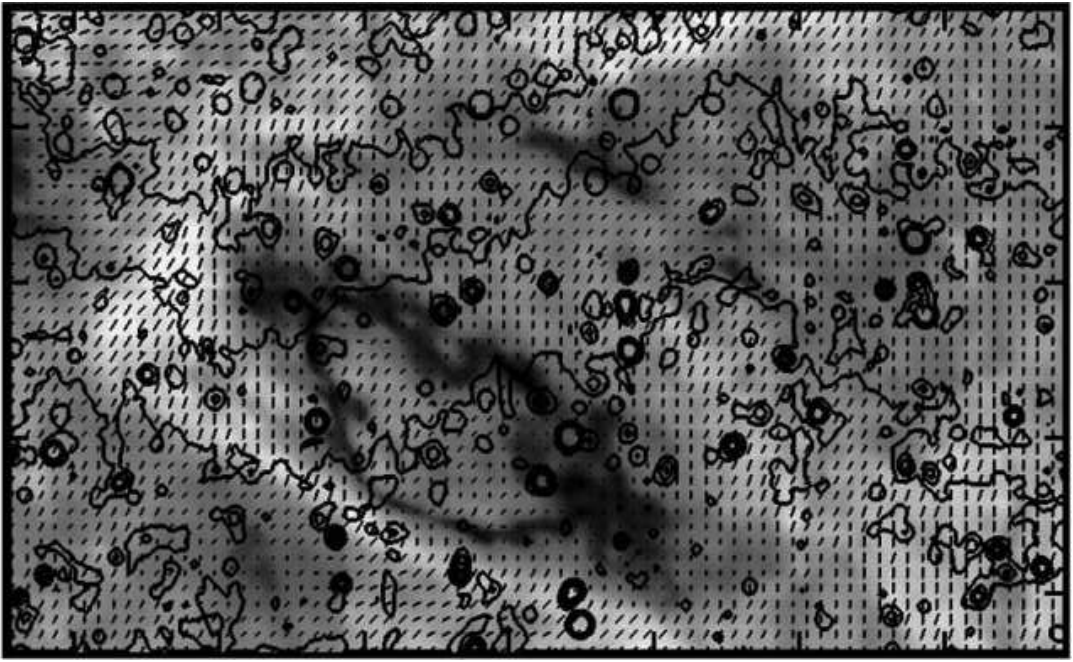}
\caption{Section of the 1.4~GHz EMLS (mapcentre at $l = 114\FDEGR1, b = -10\FDEGR6$,
mapsize $14\FDEGR6 \times 8\FDEGR3$) with large-scale
emission added from the Dwingeloo survey [45]. The map of polarized intensities
(greyscale coded) shows an elliptical shell like feature with up to $8\DEGR$ in
size. Polarization E-vectors are superimposed.
There is no counterpart in total intensities (shown by contours) indicating
systematic Faraday rotation by a shell structure of unknown distance.}
  \label{CAS}
\end{figure}

Nine Faraday screens were analysed, where the observed RM varies between $-36$~rad~m$^{-2}$
and $+26$~rad~m$^{-2}$. The typical size of the individual objects is about 2~pc.
The Faraday screens selected are seen as minima in the polarized intensity
map at 1408~MHz. At the two higher frequencies the minima are less pronounced or already
disappeared. Thus the spectral indices calculated for the minima ranges from $\beta$
= -1.3 to +2.0. The Faraday screen data were then analysed in terms of the model
shown in Fig.~\ref{model}, where a spherical cloud imposes Faraday rotation and in
addition depolarization (DP) on the background polarized emission, which than adds to
the foreground emission as observed. As a result of the model the foreground and the
background polarization is obtained and additionally the RM and the DP of the Faraday screen.
For the Taurus clouds always negative RMs were derived ranging between $-18$~rad~m$^{-2}$
and $-29.5$~rad~m$^{-2}$. This result clearly demonstrates that the RM for Faraday screens
in general is rather different to the directly observed
RM. The foreground polarization angles scatter around $0\DEGR$, while the background polarization
angles vary within $-10\DEGR$ and $-19\DEGR$.

In the case of the Taurus clouds additional information of the emissivity of the synchrotron
emission can be obtained by using the maximal fractional limit imposed by the polarized emission
to calculate the minimum total synchrotron emission. For the local Taurus clouds
enhanced synchrotron emission in this direction results, which is in agreement with other
estimates [89].

Another elliptically shaped Faraday screen from the EMLS is shown in Fig.~\ref{CAS}, which
imposes more complex although systematic effects on its background emission. The polarized
emission seen towards its centre seems basically undisturbed (compared to the surroundings
of the Faraday screen), for larger radii first a minimum is seen, followed by a
maximum in polarized emission. This clearly indicates systematic variations of the
Faraday rotation by some shell like structure, so that the rotated background emission
enhances or reduces the observed polarization when added with the foreground.
So far no distance information is available for this Faraday screen, which is required
to estimate its physical parameters.

A huge magneto-ionic bubble acting as a Faraday Screen was recently reported by
Kothes et al. [91] using for the first time combined polarization
data from the DRAO 26-m telescope, the Effelsberg 100-m telescope and the DRAO
synthesis telescope. The preliminary analysis of this shell in the Galactic anti-centre
gives a diameter of about 400~pc at a distance of 2~kpc. A HI-shell coincides with the
bubble and locates it in the Perseus arm. The origin of the bubble is likely from a stellar
wind. The object can be modelled by the following parameters: shell thickness 40\%, electron
density 0.07~cm$^{-3}$ and a magnetic field strength of 16~$\mu$G.

The problem when analysing Faraday screens from data not at an absolutely calibrated
level may be illustrated by the case of G91.8-2.5. Data from the DRAO synthesis
telescope give a $\delta$RM of about 40~rad~m$^{-2}$ measured around 1.4~GHz [92],
while RMs from Effelsberg multi-channel observations in the same
frequency range give RM = $-27$~rad~m$^{-2}$ [93]. This makes a polarization angle difference
of more than $30\DEGR$ and reflects a different amount of large-scale structures
missing in the synthesis telescope data and
the single-dish map. In both cases it seems problematic to use the measured RM values
to derive physical parameters for the Faraday screen. Its distance is needed and some
modelling is required.

\begin{figure}
\centering
\includegraphics[angle=270,bb=60 60 552 759,width=11cm,clip]{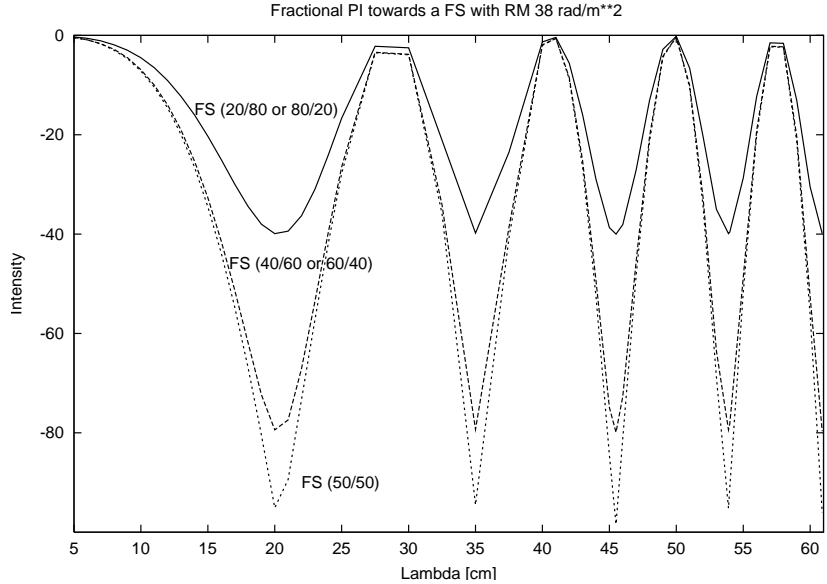}
\caption{Fractional polarization (relative to its surroundings) seen
towards a Faraday screen with a RM of 38~rad~m$^{-2}$ as a function of wavelength.
The interstellar medium along the line of sight is assumed to be
homogenous with RM = 0~rad~m$^{-2}$. The fractional polarization is
calculated for different ratios for the foreground to background emission.
In this quite simple case absolute intensity minima are expected when
 foreground and background emission relative to the Faraday screen are the same.}
  \label{model2}
\end{figure}

Low frequency polarization spectra were modelled by Vinyajkin [94]
for selected regions of the Galaxy based on an assumed multi-layer structure of
the magnetized interstellar medium. Vinyajkin's model is able to describe the observed
absorption dips at specific frequencies, which are superimposed on the power law
spectrum of the polarized emission and complements effects caused by Faraday screens.
At low frequencies small RM variations already cause large changes on the polarized
structures, while at higher frequencies larger RMs as revealed by discrete Faraday screens
are needed for observable polarization effects.

Figure~\ref{model2} shows an example of the
influence of a discrete Faraday screen on the observed polarized intensities. In this case
the RM of the interstellar medium is assumed to have RM = 0~rad~m$^{-2}$ along
the entire line of sight and thus the Faraday screen always causes a depression,
whose amount depends on its location within the diffuse interstellar medium. The
minimum polarized intensity is observed for equal foreground and background
emission. It is easily seen from  Fig.~\ref{model2} that Faraday screens of different RM
located at different distances in the interstellar medium are able to
create a rather complex distribution of polarized intensities even though the
diffuse magneto-ionic interstellar medium has very uniform properties. It is also
evident that at higher frequencies the Faraday screens need to have large RMs to
cause an effect on the polarized distribution. High frequency polarization observations
are needed to reveal the intrinsic properties of the magnetic field distribution in
the interstellar medium.

Optically thin HII-regions may act as Faraday screens by depolarizing
their background polarization.
They become optically thick at low frequencies, depending on their electron density
and temperature. Because their distance can be independently determined, they are ideal
objects for 'Galactic tomography'. At low frequency the foreground total intensity can be
measured as they fully absorb the background emission. At sufficiently high frequencies
(depending on their distance) the corresponding foreground
polarization component is observed. Thus future high angular resolution low frequency
observations, as they will become possible with the LOFAR telescope, when
combined with high frequency polarization observations, e.g. from the Effelsberg 100-m
telescope, will provide detailed information on the magnetic field components within a
few kpc and the distribution of the thermal gas components [95]. Low frequency
absorption spectra need some modelling of the components of the thermal gas along
the line of sight [96]. These results will constrain Faraday screen models describing
the radiation transfer of polarized emission.

\subsection{Canals}

May be the most striking unusual structures in the various new polarization survey maps
with sufficiently high angular resolution are long narrow nearly depolarized features commonly
named 'canals'. They are for example clearly visible in Fig.~\ref{anti1} or Fig.~\ref{anti2}.
Most of these 'canals' have similar properties: their width is about one beam and the
polarization angles change from one side of the 'canal' to the other by
$90\DEGR$. This requires a gradient for the U and Q intensities perpendicular to
the 'canals'. At the 'canal' itself U and Q become zero.

So far two explanations have been published on the origin of the 'canals': beam depolarization
[54, 55] caused by a sufficiently large RM gradient within one beam and Faraday depth
depolarization [97] in a uniform synchrotron emitting magneto-ionic medium. In the case
of beam depolarization the 'canals' do not change their position with frequency, but
at high frequencies 'canals' require very large RM changes across a beam. At 350~MHz a RM
gradient across
a beam of about 2.1~rad~m$^{-2}$ is sufficient and no unusual properties
of the interstellar medium are needed. Some numerical models of a Faraday rotating interstellar
medium supports this interpretation of low frequency observations [55].
At 4.8~GHz, however, the RM gradient across a beam width must be
400~rad~m$^{-2}$ to cause a 'canal', which seems to be an unrealistic high value
for the interstellar medium in general and therefore
'canals' observed at high frequencies most likely need another explanation. Shukurov and
Berkhuijsen [97] propose differential Faraday rotation as an explanation for the 'canals':
The 'canals'
represent 'level lines' of RM. Because they are no physical structures they have been called
'Faraday ghosts'. The separation of the 'canals', however, provides useful information on the
turbulent interstellar medium.

It must be noted that the discussion of 'canals' above is in all
cases based on relative (synthesis telescope) rather than absolute polarization data.
Shukurov and Berkhuijsen [97] comment on various aspects of that problem. Obviously
some conclusions reached for the 'canals'
might change when the large-scale polarized emission is available and has been properly taken
into account.

\subsection{RM data from pulsars and extragalactic sources}

RM data from pulsars and extragalactic sources are potentially very valuable
for the interpretation of diffuse Galactic polarized
emission as they trace the entire line of sight component of the Galactic magnetic
field in a certain direction. They are thus complementary to the perpendicular
component of the magnetic field as seen from synchrotron emission. However, tomography
is needed to reveal information on the magnetic field and the thermal electron density
along the line of sight. Figure~\ref{rm-titus}
illustrates the relation of RMs of extragalactic sources in a certain area to the RMs
measured for the diffuse polarized Galactic emission.

In particular RM data from pulsars are useful to analyse the Galactic magnetic field as
they provide in principle in combination with the measurable dispersion measure (DM)
the line of sight magnetic field strength (however see [98] for related problems).
While the RM of a single pulsar is helpful, for instance, to estimate
the foreground effect of the interstellar medium towards polarized SNRs,
fairly large RM samples are needed for a reliable estimate of the local
structure, direction and intensity of the magnetic field. Current estimates
give a regular magnetic field strength around 1.4~$\mu$G in the direction
of the local arm with a pitch angle of a few degrees [27, 28, 29, 30], which
is small when compared to the total magnetic field strength (see Sect. 4.2). Also
magnetic field reversals between different spiral arms of the Galaxy are based
on pulsar RM as discussed in Sect. 4.2.

There are various observation projects actually carried out aiming to increase
the number of pulsar RMs and the RMs of extragalactic sources. In particular from
synthesis telescope surveys, where narrow band polarimetry is available, a large number of
RMs of sources located in the Galactic plane were measured in the course of the
Canadian Galactic Plane Survey (CGPS) [68] and the Southern Galactic Plane Survey
(SGPS) [81] (see Sect. 6.5). These densely sampled RM data of extragalactic sources
are obtained at 1.4~GHz with a beam of about 1$\arcmin$ [82, 99]. At higher latitudes
single-dish telescopes are used to increase the number of
RMs of extragalactic sources. A project was recently completed to measure 1800 polarized
NVSS sources with the Effelsberg 100-m telescope mostly located out of the
Galactic plane using multi-channel polarimetry at 21cm/18cm [Han et al., in prep.].

The potential of a very dense grid of RM data from pulsars and extragalactic sources
for a detailed analysis of the properties of the Galactic magnetic field on all relevant
scales and also for the evolution of magnetic fields in galaxies is widely accepted.
Consequently one of the key-projects for the planned Square Kilometer Array (SKA) is
a RM survey providing RM data for some tens of million sources [100]. This RM survey
is planned as part of a global sky survey in the 1.4~GHz range. Based on model predictions
about 2900 polarized sources per deg$^{2}$ stronger than 1~$\mu$Jy are expected and
about 50\% of them should have a measurable RM. This gives about 2~10$^{7}$~RMs from the
survey with a mean distance of about $90\arcsec$ between the sources. This holds for a 1~h
integration time with the SKA. An increase of the integration time to 10~h will reduce the mean
RM spacing in that field to $40\arcsec$ [100]. Details on numerous important scientific
applications of such a dense RM grid were discussed by Beck and Gaensler [100].

\section{Status and Outlook}

The revival of Galactic polarization surveys over the last years has led to intensive
discussions on the composition and structure of the Galactic magneto-ionic medium.
Numerous workshops and conferences were held during the last years and the proceedings
reflect the observational status as well as the progress achieved in modelling the
magneto-ionic interstellar medium [101-104].

Many large polarization surveys are in progress today mostly around 1.4~GHz.
The direction of future observations is towards higher frequencies, where local Faraday
rotation effects are less important and the Galaxy becomes transparent even in the
disk. Such measurements require high
sensitivity equipment because of the weakness of the polarized signals.
Multi-channel polarization observations in numerous narrow bands are another
observational direction, which in combination with advanced analysis methods [38]
is quite powerful to decompose
complex superimposing polarization structures along the line of sight. The analysis
of the planned L-band RM-survey with the DRAO 26-m antenna (see Sect. 6.5) relies
on this technique.

Absolutely calibrated polarization data at high frequencies (above 1.4~GHz) for
large-scale emission is another indispensable need for the proper analysis of polarization data.
Unfortunately such data are not easy to obtain at high frequencies. Beside sensitive
receivers also a very high stability of the entire receiving system is needed as well as
a low level of the telescope's far-sidelobes. Such measurements are time consuming,
but can be done with small telescopes. However, they must be located at well selected
suitable sites.

Galactic polarization is considered as an important foreground for future sensitive
Cosmic Microwave Background (CMB) polarization observations. Polarization
surveys available in numerical form as the Leiden-Dwingeloo large-scale survey
were analysed in terms of
their angular power spectrum by various authors [105 and references therein].
Also the new large-scale polarization surveys are actually analysed in a similar
way [106]. Of course, also for the various recent Galactic plane surveys
with higher angular resolution the angular power spectrum is available although they
are more influenced by Faraday effects than high latitude regions [107 and
references therein].
A problem with available Galactic polarization data are their low frequencies and the
influence of Faraday effects when compared to the much higher frequencies, where
CMB observations are carried out. Source contamination of the angular power spectrum
turns out to be severe, depolarization effects vary with frequency and only at high
Galactic latitudes one can be sure to look out off the Galaxy. Thus extrapolations towards
higher frequencies are not easy to perform with high accuracy, but upper limits can be
given with confidence.

High frequency observations above 1.4~GHz are needed. 2.7~GHz and 5~GHz observations are
already on the way, which may better constrain foreground predictions at CMB frequencies in general
than it is possible with available data. However, recent results for selected fields with low
synchrotron and dust emission are promising. CMB Polarization (CMBP) is usually expressed in terms
of E- and B-modes [e.g. 108], where the E-mode is at a few percent of the CMB anisotropies, but the
level of the much weaker B-mode is largely unknown. However, the B-mode is needed to disentangle
different Inflation models [109]. Sensitive observations of a northern and southern target selected
for CMB Polarization (CMBP) observations [110] by the BaR-SPOrt experiment [111]
were recently carried out. 1.4~GHz observations with the Effelsberg 100-m telescope of the
northern BaR-SPOrt field, which is at very
high Galactic latitude, reveals the lowest contamination by polarized Galactic foreground emission
measured so far. This implies good chances to detect B-mode emission at about 90~GHz
[112]. More extended searches may reveal areas with are suited to detect even weaker
B-mode signals. Therefore the polarized Galactic synchrotron foreground seems not to be a
limitation to detect weak CMBP signals at least on scales up to a few degrees in extent.

\section{Acknowledgement}

I like to thank Ernst F\"urst and Patricia Reich for sharing common efforts over many years
on Galactic survey projects and their critical comments on the manuscript. I acknowledge
discussions and cooperation with Roy Duncan, JinLin Han, Roland Kothes, Tom Landecker
and Rob Reid on polarization survey issues. I am grateful to Richard Wielebinski for
initiating and strongly supporting the various MPIfR survey projects over the last 30~years.
Without his engagement the new 5~GHz polarization surveys would not
been possible. Former PhD-students made significant contributions to the different
survey projects at the MPIfR in Bonn: Norbert Junkes, Juan Carlos Testori,
Buelent Uyan{\i}ker and Maik Wolleben. Helpful discussions with Rainer Beck on the manuscript are acknowledged too.

\section{References}

\begin{enumerate}

\item C.H. Mayer, T.P. McCullough, and R.M. Sloanaker, 1957, Astrophys. J., 126, 468

\item G. Westerhout, C.L. Seeger, W.N. Brouw, and J. Tinbergen, 1962, Bull. Astron. Inst.
Netherlands, 16, 187

\item R. Wielebinski, J.R. Shakeshaft, and I.I.K. Pauliny-Toth, 1962, The Observatory, 82, 158

\item R. Wielebinski, and J.R. Shakeshaft, 1964, MNRAS, 128, 19

\item E.M. Berkhuijsen, and W.N. Brouw, 1963, Bull. Astron. Inst. Netherlands, 17, 185

\item D.S. Mathewson, N.W. Broten, and D.J. Cole, 1965, Austr. Journal of Physics, 18, 665

\item C.A. Muller, E.M. Berkhuijsen, W.N. Brouw, and J. Tinbergen, 1963, Nature, 200, 155

\item E.M. Berkhuijsen, W.N. Brouw, C.A. Muller, and J. Tinbergen, 1964, Bull. Astron. Inst. Netherlands, 17, 465

\item R.G. Bingham, 1966, MNRAS, 134, 327

\item D.S. Mathewson, N.W. Broten, and D.J. Cole, 1966, Austr. Journal of Physics, 19, 93

\item E.M. Berkhuijsen, 1975, Astron. Astrophys., 40, 311

\item A.G. Pacholcyzk, 1970, Radio Astrophysics, San Francisco, W.H. Freemann

\item P. Reich, 2003, Acta Astronomica Sinica, Vol. 44 Suppl., 130

\item J.L. Jonas, E.E. Baart, and G.D. Nicolson, 1998, MNRAS, 297, 977

\item C.L. Bennett, M. Halpern, G. Hinshaw, et al., 2003, ApJS, 148, 1

\item R. Wielebinski, 2005, in 'Cosmic Magnetic Fields', Eds. R. Wielebinski, R. Beck,
Lect. Notes Phys. 664, Springer, 89

\item W. Reich, 1982,  Astron. Astrophys. Suppl., 48, 219

\item P. Reich, and W. Reich, 1986,  Astron. Astrophys. Suppl., 63, 205

\item P. Reich, J.C. Testori, and W. Reich, 2001,  Astron. Astrophys., 376, 861

\item P. Reich, W. Reich, and J.C. Testori, 2004,  in 'The Magnetized Interstellar Medium',
Eds. B. Uyan{\i}ker, W. Reich, R. Wielebinski, Copernicus GmbH, 63

\item R. Beck, 2005, in 'Cosmic Magnetic Fields', Eds. R. Wielebinski, R. Beck,
Lect. Notes Phys. 664, Springer, 41

\item M. Krause, 2004,  in 'The Magnetized Interstellar Medium', Eds.
B. Uyan{\i}ker, W. Reich, R. Wielebinski, Copernicus GmbH, 173

\item K. Beuermann, G. Kanbach, and E.M. Berkhuijsen, 1985, Astron. Astrophys., 153, 17

\item S. Phillipps, S. Kearsey, J.L. Osborne, C.G.T. Haslam, and H. Stoffel, 1981, Astron.
Astrophys., 98, 286

\item C.G.T. Haslam, H. Stoffel, C.J. Salter, and W.E. Wilson, 1982, Astron. Astrophys. Suppl., 47, 1

\item A.W. Strong, I.V. Moskalenko, and O. Reimer, 2000, ApJ, 537, 763

\item R.J. Rand, and A.G. Lyne, 1994, MNRAS, 268, 497

\item J.L. Han, and G.J. Qiao, 1994, Astron. Astrophys., 288, 759

\item C. Indrani, and A.A. Deshpande, 1999, New Astronomy, 4, 33

\item J.L. Han, R.N. Manchester, A.G. Lyne, and G.J. Qiao, 2002, ApJ, 570, L17

\item A. Shukurov, 2005, in 'Cosmic Magnetic Fields', Eds. R. Wielebinski, R. Beck,
Lect. Notes Phys. 664, Springer, 113

\item J.H. Taylor, and J.M. Cordes, 1993, ApJ, 411, 674

\item J.M. Cordes, and T.J.W. Lazio, 2005, astro-ph$\backslash$0207156

\item C.K. Lacey, T.J.W. Lazio, N.E. Kassim, N. Duric, D.S. Briggs, and K.K. Dyer, 2001, ApJ 559, 954

\item K.R. Anantharamaiah, 1986, J. Astrophys. Astron., 7, 131

\item R.J. Reynolds, L.M. Haffner, and S.L. Tufte, 1999, ApJ, 525, L21

\item R.-J. Dettmar, 2003, Acta Astronomica Sinica, Vol. 44 Suppl., 106

\item M.A. Brentjens, and A.G. de Bruyn, 2005, Astron. Astrophys., 441, 1217

\item A.G. de Bruyn, and M.A. Brentjens, 2005, Astron. Astrophys., 441, 931

\item B.J. Burn, 1966, MNRAS, 133, 67

\item E.N. Vinyajkin, and V.A. Razin, 2002, in 'Astrophysical Polarized Backgrounds', Eds. S. Cecchini
et al., AIP 609, 26

\item A. Wilkinson, F.G. Smith, 1974, MNRAS, 167, 593

\item M. Wolleben, 2005, PhD-Thesis, Bonn University

\item P.M.W. Kalberla, U. Mebold, and W. Reich, 1980, Astron. Astrophys., 82, 275

\item W.N. Brouw, and T.A.Th. Spoelstra, 1976, Astron. Astrophys. Suppl., 26, 129

\item T.A.Th. Spoelstra, 1984, Astron. Astrophys., 135, 238

\item M.H. Wieringa, A.G. de Bruyn, D. Jansen, W.N. Brouw, and P. Katgert, 1993,
Astron. Astrophys., 268, 215

\item M. Haverkorn, P. Katgert, and A.G. de Bruyn, 2000, Astron. Astrophys., 356, L13

\item M. Haverkorn, P. Katgert, and A.G. de Bruyn, 2002, in 'Astrophysical Polarized Backgrounds',
Eds. S. Cecchini et al., AIP 609, 72

\item M. Haverkorn, P. Katgert, and A.G. de Bruyn, 2003, Astron. Astrophys., 403, 1031

\item M. Haverkorn, P. Katgert, and A.G. de Bruyn, 2003, Astron. Astrophys., 403, 1045

\item M. Haverkorn, P. Katgert, and A.G. de Bruyn, 2003, Astron. Astrophys., 404, 233

\item M. Haverkorn, P. Katgert, A.G. de Bruyn, and F. Heitsch, 2004, in 'The Magnetized Interstellar
 Medium', Eds. B. Uyan{\i}ker, W. Reich, R. Wielebinski, Copernicus GmbH, 81

\item M. Haverkorn, P. Katgert, and A.G. de Bruyn, 2004, Astron. Astrophys. 427, 549

\item M. Haverkorn, and F. Heitsch, 2004, Astron. Astrophys., 421, 1011

\item M. Wolleben, T.L. Landecker, W. Reich, and R. Wielebinski, 2006,
Astron. Astrophys., in press

\item M. Wolleben, T.L. Landecker, W. Reich, and R. Wielebinski, 2004, in
'The Magnetized Interstellar Medium', Eds. B. Uyan{\i}ker, W. Reich, R. Wielebinski,
Copernicus GmbH, 51

\item J.C. Testori, P. Reich, and W. Reich, 2004,  in 'The Magnetized Interstellar Medium',
Eds. B. Uyan{\i}ker, W. Reich, R. Wielebinski, Copernicus GmbH, 57

\item W. Reich, P. Reich, and E. F\"urst, 1990, Astron. Astrophys. Suppl., 83, 539

\item P. Reich, W. Reich, and E. F\"urst, 1997, Astron. Astrophys. Suppl., 126, 413

\item W. Reich, E. F\"urst, C.G.T. Haslam, P. Steffen, and K. Reif, 1984,
Astron. Astrophys. Suppl., 58, 197

\item W. Reich, E. F\"urst, P. Reich, and K. Reif, 1990, Astron. Astrophys. Suppl., 85, 633

\item E. F\"urst, W. Reich. P. Reich, and K. Reif, 1990, Astron. Astrophys. Suppl., 85, 691

\item W.J. Altenhoff, D. Downes, T. Pauls, and J. Schraml, 1979, Astron. Astrophys. Suppl., 35, 23

\item R.F. Haynes, J.L. Caswell, and L.W.J. Simons, 1978, Aust. J. of Physics, Astrophysical
Suppl., 45, 1

\item A.R. Duncan, R.T. Stewart, R.F. Haynes, and K.L. Jones, 1995, MNRAS, 277, 36

\item T. Handa, Y. Sofue, N. Nakai, H. Hirabayashi, and M. Inoue, 1987, PASJ, 39, 709

\item A.R. Taylor, S.J. Gibson, M. Peracaula, et al., 2003, AJ, 125, 3145

\item N. Junkes, E. F\"urst, and W. Reich, 1987, Astron. Astrophys. Suppl., 69, 451

\item A.R. Duncan, P. Reich, W. Reich, and E. F\"urst, 1999, Astron. Astrophys., 350, 447

\item A.R. Duncan, R.F. Haynes, K.L.. Jones, and R.T. Stewart, 1997, MNRAS, 291, 279

\item W. Reich, E. F\"urst, P. Reich, B. Uyan{\i}ker, R. Wielebinski, and M. Wolleben, 2004, in
'The Magnetized Interstellar Medium', Eds. B. Uyan{\i}ker,
W. Reich, R. Wielebinski, Copernicus GmbH, 45

\item B. Uyan{\i}ker, E. F\"urst, W. Reich, P. Reich, and R. Wielebinski, 1998,
Astron. Astrophys. Suppl., 132, 401

\item B. Uyan{\i}ker, E. F\"urst, W. Reich, P. Reich, and R. Wielebinski, 1999,
Astron. Astrophys. Suppl., 138, 31

\item Z.Z. Abidin, J.P. Leahy, A. Wilkinson, P. Reich, W. Reich, and R. Wielebinski, 2003,
New Astronomy Review, Vol.47, Issue 11-12, 1151

\item  The GALFA Consortium, 2003, Galactic Astronomy with the Arecibo L-band feed array (ALFA)
(http://alfa.naic.edu/galactic/alfa\_galactic.html)

\item B. Uyan{\i}ker, T.L. Landecker, A.D. Gray, and R. Kothes, 2003, ApJ, 585, 785

\item T.L. Landecker, B. Uyan{\i}ker, and R. Kothes, 2002, in 'Astrophysical
Polarized Backgrounds', Eds. S. Cecchini et al., AIP 609, 9

\item R. Kothes, and T.L. Landecker, 2004, in 'The Magnetized Interstellar Medium',
Eds. B. Uyan{\i}ker, W. Reich, R. Wielebinski, Copernicus GmbH, 33

\item R.I. Reid, 2004, in 'The Magnetized Interstellar Medium',
Eds. B. Uyan{\i}ker, W. Reich, R. Wielebinski, Copernicus GmbH, 39

\item B.M. Gaensler, J.M. Dickey, N.M. McClure-Griffiths, A.J. Green, M.H. Wieringa,
and R.F. Haynes, 2001, ApJ, 549, 959

\item M. Haverkorn, B.M. Gaensler, J.C. Brown, N.M. McClure-Griffiths, J.M. Dickey, and A.J. Green,
2005, astro-ph$\backslash$0511407

\item X.H. Sun, W. Reich, J.L. Han, P. Reich, and R. Wielebinski, 2006, Astron. Astrophys., in press.

\item S. Rys, K.T. Chyzy, A. Kus, E. Pazderski, M. Soida, and M. Urbanik, 2006, AN, in press

\item W. Reich, E. F\"urst, P. Reich, R. Wielebinski, and M. Wolleben, 2002, in 'Astrophysical
Polarized Backgrounds', Eds. S. Cecchini et al., AIP 609, 3

\item R. Reid, 2005, CGPS News Vol. 33

\item D.D. Sokoloff, A.A. Bykov, A. Shukurov, E.M. Berkhuijsen, R. Beck, and A.D. Poezd, 1998, MNRAS,
299, 189

\item A.D. Gray, T.L. Landecker, P.E. Dewdney, A.R. Taylor, A.G. Willis, and M. Normandeau,
1999, ApJ, 514, 221

\item M. Wolleben, and W. Reich, 2004, in 'The Magnetized Interstellar Medium',
Eds. B. Uyan{\i}ker, W. Reich, R. Wielebinski, Copernicus GmbH, 99

\item M. Wolleben, and W. Reich, 2004, Astron. Astrophys., 427, 537

\item R. Kothes, T.L. Landecker, M. Wolleben, T. Foster, and W. Reich, 2004, CGPS News, Vol. 31

\item B. Uyan{\i}ker, and T.L. Landecker, 2002, ApJ, 575, 225

\item B. Uyan{\i}ker, 2004, in 'The Magnetized Interstellar Medium',
Eds. B. Uyan{\i}ker,
W. Reich, R. Wielebinski, Copernicus GmbH, 71

\item E.N. Vinyajkin, 2004, in 'The Magnetized Interstellar Medium',
Eds. B. Uyan{\i}ker, W. Reich, R. Wielebinski, Copernicus GmbH, 93

\item W. Reich, 2005, AN, 326, 620

\item J.D. Peterson, and W.R. Webber, 2002, ApJ, 575, 217

\item A. Shukurov, and E.M. Berkhuijsen, 2003, MNRAS, 342, 496

\item R. Beck, A. Shukurov, D. Sokoloff, and R. Wielebinski, 2003, Astron. Astrophys., 411, 99

\item J.C. Brown, A.R. Taylor, and B.J. Jackel, 2003, ApJS, 145, 213

\item R. Beck, and G.M. Gaensler, 2004, in 'Science with the Square Kilometre Array', Eds.
C. Carilli, S. Rawlings, New Astronomy Reviews, 48, 1289

\item E.M. Berkhuijsen (Ed.), 1999, 'Galactic Foreground Polarization',
Max-Planck-Institut f\"ur Radioastronomie

\item T.L. Landecker (Ed.), 2001, 'Radio Polarization: A New Probe of the Galaxy', DRAO
Penticton

\item S. Cecchini, S. Cortiglioni, R. Sault, and C. Sbarra (Eds.), 2002, 'Astrophysical
Polarized Backgrounds', AIP, Vol. 609

\item B. Uyan{\i}ker, W. Reich, and R. Wielebinski (Eds.), 2004, 'The
Magnetized Interstellar Medium', Copernicus GmbH  \\
 (http://www.mpifr-bonn.mpg.de/div/konti/antalya/contrib.html)

\item L. La Porta, and C. Burigana, 2006, Astron. Astrophys., in press

\item C. Burigana, L. La Porta, P. Reich, and W. Reich, 2006, Astron. Nachrichten, in press

\item M. Tucci, E. Carretti, S. Cecchini, L. Nicastro, R. Fabbri, B.M. Gaensler,
J.M. Dickey, and N.M. McClure-Griffiths, 2002, ApJ, 579, 607

\item M. Zaldarriaga, and U. Seljak, 1997, Phys. Rev. D., 55, 1830

\item W.H. Kinney, 1998, Phys. Rev. D., 58, 123506

\item E. Carretti, G. Bernardi, S. Cecchini, et al., 2002, in 'Experimental Cosmology at Millimetre
 Wavelength', Eds. M. De Petri, M. Gervasi, AIP 616, 140

\item S. Cortiglioni, et al., 2003, in '16th ESA Symposium on European Rocket and Balloon
Programmes and Related Research', Ed. B. Warmbein, ESA Proc. SP-530, 271

\item E. Carretti, S. Poppi, W. Reich, P. Reich, E. F\"urst, G. Bernardi, S. Cortiglioni,
and C. Sbarra, 2006, MNRAS, in press

\end{enumerate}

\end{document}